\documentclass[a4paper,fleqn,usenatbib]{mnras}

\usepackage{newtxtext,newtxmath}

\usepackage[T1]{fontenc}
\usepackage{ae,aecompl}

\usepackage{graphicx}	
\usepackage{amsmath}	
\usepackage{amssymb}	
\usepackage{multirow}
\usepackage{epsfig}
\usepackage{subfigure}

\title[The lively accretion disk in NGC 2992. Transient iron K emission lines.]{The lively accretion disk in NGC 2992. I. Transient iron K emission lines in the high flux state}

\author[Andrea Marinucci, et al.]{A. Marinucci$^{1}$\thanks{E-mail: andrea.marinucci@asi.it (AM)}, S. Bianchi$^{2}$, V. Braito$^{3,4}$, B. De Marco$^{5}$, G. Matt$^{2}$, R. Middei$^{6,7}$, 
\newauthor E. Nardini$^{8,9}$, J. N. Reeves$^{3,4}$\\\\
$^1$ASI - Unit\`a di Ricerca Scientifica, Via del Politecnico snc, 00133, Roma, Italy\\
$^2$Dipartimento di Matematica e Fisica, Universit\`a degli Studi Roma Tre, via della Vasca Navale 84, 00146 Roma, Italy\\
$^3$Center for Space Science and Technology, University of Maryland Baltimore County, 1000 Hilltop Circle, Baltimore, MD 21250, USA\\
$^4$INAF - Osservatorio Astronomico di Brera, Via Bianchi 46 I-23807 Merate (LC), Italy\\
$^5$Nicolaus Copernicus Astronomical Center, Polish Academy of Sciences, Bartycka 18, PL-00-716 Warsaw, Poland\\
$^6$INAF - Osservatorio Astronomico di Roma, Via Frascati 33, 00078, Monte Porzio Catone (Roma), Italy\\
$^7$Space Science Data Center - ASI, Via del Politecnico s.n.c., 00133 Roma, Italy\\
$^8$Dipartimento di Fisica e Astronomia, Universit\`a di Firenze, via G. Sansone 1, I-50019 Sesto Fiorentino, Firenze, Italy\\
$^9$INAF - Osservatorio Astrofisico di Arcetri, Largo Enrico Fermi 5, 50125 Firenze, Italy\\
}

\date{Accepted XXX. Received YYY; in original form ZZZ}

\pubyear{2020}

\begin{document}
\label{firstpage}
\pagerange{\pageref{firstpage}--\pageref{lastpage}}
\maketitle

\begin{abstract}
We report on one of the brightest flux levels of the Seyfert 2 galaxy NGC 2992 ever observed in X-rays, on May 2019. The source has been monitored every few days from March 26, 2019 to December 14, 2019 by {\it Swift}-XRT, and simultaneous XMM-{\it Newton} (250 ks) and {\it NuSTAR} (120 ks) observations were triggered on May 6, 2019. The high count rate of the source (its 2-10 keV flux ranged between 0.7 and $1.0\times10^{-10}$ erg cm$^{-2}$ s$^{-1}$) allows us to perform a time-resolved spectroscopy, probing spatial scales of tens of gravitational radii from the central black hole. By constructing a map of the excess emission over the primary continuum, we find several emission structures in the 5.0-7.2 keV energy band. From fitting the 50 EPIC pn spectral slices of $\sim$5 ks duration, we interpret them as a constant narrow iron K$\alpha$ line and three variable components in the iron K complex. When a self-consistent model accounting for the accretion disk emission is considered ({\sc KYNrline}), two of these features (in the 5.0-5.8 keV and 6.8-7.2 keV bands) can be ascribed to a flaring region of the accretion disk located at ${r_{in}}\simeq15$-40 r$_{g\rm }$ from the black hole. The third one (6.5-6.8 keV) is likely produced at much larger radii ($r_{in}>50$ r$_{g\rm }$). The inner radius and the azimuthal extension retrieved from the coadded spectra of the flaring states are ${ r_{in}}=15\pm3$ r$_{g\rm }$ and $\phi=165^{\circ}-330^{\circ}$, suggesting that the emitting region responsible for the broad iron K component is a relatively compact annular sector within the disk. Our findings support a physical scenario in which the accretion disk in NGC 2992 becomes more active at high accretion rates ($L_{\rm bol}/L_{\rm Edd}\geq4\%$).

\end{abstract}

\begin{keywords}
Galaxies: active - Galaxies: Seyfert - Galaxies: accretion - Individual: NGC 2992
\end{keywords}



\section{Introduction}
Active Galactic Nuclei (AGN) that show high X-ray variability on relatively short (less 
than hundreds of ks) timescales are the perfect astrophysical laboratories for studying the response of the accretion disk to changes of the primary continuum. Spectral features both in emission and in absorption have been detected in bright sources and can be studied to infer the physical properties of the circumnuclear matter \citep[MCG-6-30-15, IRAS 13224-3809:][]{ify99,fv03,mmm14,ppf17,pab17}. On the other hand, the reverberation of the X-ray radiation reprocessed by the accretion disk \citep{demarco13,ucf14,kaf16} suggests that the hot corona, responsible for the primary continuum, has a typical size of a few gravitational radii and is located close to the central supermassive black hole (SMBH). This is also supported by microlensing experiments on quasars \citep{ckd09,gds17}.\\
A narrow neutral iron K$\alpha$ emission line at 6.4 keV is ubiquitous in nearby AGN \citep{george00,per02,bianchi07} and when a broad component is detected, this is indicative of special and general relativistic effects \citep{nan97,fab00,rn03}  occurring at a few gravitational radii r$_g=GM/c^2$ from the central SMBH. Many studies on the variability pattern of such a broad iron K$\alpha$ component have been presented in the past \citep{imf04, lnp04,tmg06,ppm07,tdm07,dmi09,npr16} , typically on time scales of tens of ks.

NGC 2992 is a nearby \citep[z=0.00771:][]{keel96} Seyfert 1.9/1.5 galaxy \citep{trippe08}. In X-rays, it is absorbed by a column density $N_{\rm H} \sim9\times10^{21}$ cm$^{-2}$ and it has been observed to vary in flux by up to a factor of 20 in a few years, and by almost an order of magnitude on time scales of days \citep[0.8-8.9$\times10^{-11}$ erg cm$^{-2}$ s$^{-1}$:][]{mkt07}. Even though the source has been extensively observed by all major X-ray satellites, its high flux level is still poorly studied. The time variability of the iron K$\alpha$ line is intriguing, suggesting the presence of a broad iron K$\alpha$ component, between 5 and 6 keV, which becomes more intense at high flux levels \citep{ymg07,sym10,mbb18}.
This is the opposite of what is usually observed in other bright sources with relativistic lines and explained in the framework of the light bending model \citep{mama96,mf04}.\\
\indent We hereby present results from the XMM-{\it Newton} and {\it NuSTAR} observations of a high flux level of NGC 2992 occurred in May 2019 (F$_{2-10}>7\times10^{-11}$ erg cm$^{-2}$ s$^{-1}$), with the aim of constraining the Fe K line emission regions, by monitoring the variability patterns of the line. The paper is structured as follows: in Sect. 2 we discuss the observations and data reduction, in Sect. 3  we present the iron K excess map, the excess variance $F_{\rm var}$ spectrum and the spectral analyses. We discuss and summarize the physical implications of our results in Sect. 4 and 5. 

\section{Observations and data reduction}
\subsection{Swift}
The past RXTE $\sim1$ year light curve \citep{mkt07} showed a very broad range of 2-10 keV flux levels for NGC 2992 (0.8-8.9$\times10^{-11}$ erg cm$^{-2}$ s$^{-1}$). Our {\it Swift}-XRT monitoring program is composed of 60 observations, $\sim$2 ks long each, from March 26, 2019 to December 14, 2019. The source was targeted every 2 days during the XMM-{\it Newton} observing windows and every 4 days in the remaining months. The requested high flux threshold (F$_{2-10}>7\times10^{-11}$ erg cm$^{-2}$ s$^{-1}$) was met on May 6, 2019, with a 2-10 keV flux F$_{2-10}$=7.0$\pm0.4\times10^{-11}$ erg cm$^{-2}$ s$^{-1}$. The complete {\it Swift} observational campaign will be analysed in a separate paper (Middei et al., in prep.).
\subsection{XMM-Newton}
XMM-{\it Newton} started its observation of NGC 2992 on May 7, 2019 for two consecutive orbits (ObsIDs  0840920201, 0840920301) with the EPIC CCD cameras, the pn \citep{struder01} and the two MOS \citep{turner01}, operated in small window and medium filter mode. Data from the MOS detectors are not included in our analysis due to pile-up. The data from the pn camera show no significant pile-up as indicated by the {\sc epatplot} output. The extraction radii and the optimal time cuts for flaring particle background were computed with SAS 18 \citep{gabr04} via an iterative process which leads to a maximization of the Signal to Noise Ratio (SNR), similar to the approach described in \citet{pico04}. The resulting optimal extraction radii for the source and the background spectra are 40 and 50 arcsec, respectively. 
The net exposure times for the two time-averaged spectra are and 92.6 and 92.8 ks. 
We then decided to extract spectra from 5.8 ks time intervals during the XMM-{\it NuSTAR} simultaneous observations (191-301 ks from the beginning of the XMM pointing) and from 5 ks time intervals when XMM data alone are available. This choice allows us to have a regular time spacing which is centered on the {\it NuSTAR} on-source pointing. With this choice, we obtained 26 and 24 spectra for the first and second XMM orbits, respectively.  Spectra were then binned in order to over-sample the instrumental resolution by at least a factor of three and to have no less than 30 counts in each background-subtracted spectral channel, for the spectral fitting procedure. 
We adopt the cosmological parameters $H_0=70$ km s$^{-1}$ Mpc$^{-1}$, $\Omega_\Lambda=0.73$ and $\Omega_m=0.27$, i.e. the default ones in \textsc{Xspec 12.10.1} \citep{Xspec}. Errors correspond to the 90\% confidence level for one interesting parameter ($\Delta\chi^2=2.7$), if not stated otherwise. 
\begin{figure}
 \epsfig{file=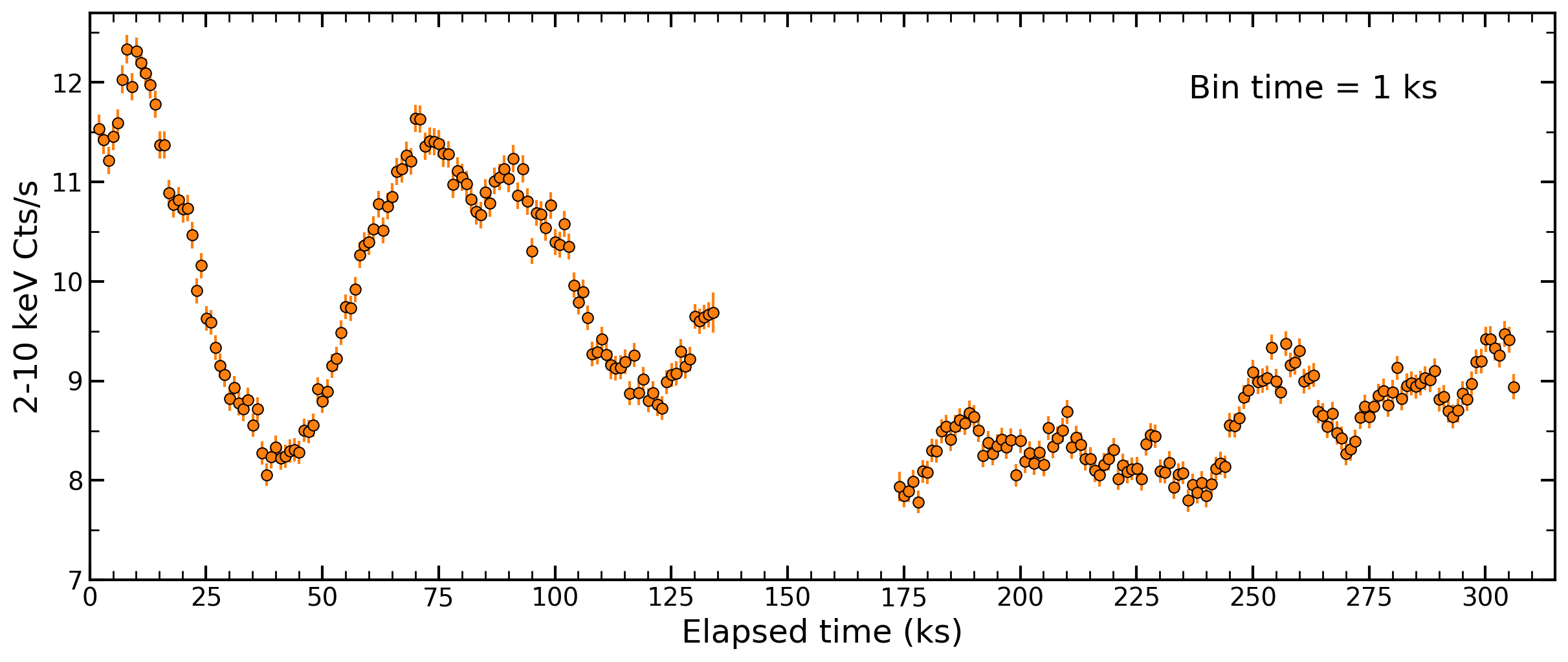, width=\columnwidth}
  \caption{EPIC pn light curve of the two consecutive observations of NGC 2992, in the 2-10 keV energy band.}
  \label{lcurve}
\end{figure}
\subsection{NuSTAR}
{\it NuSTAR} (Harrison et al. 2013) observed NGC 2992 with its two co-aligned X-ray telescopes with corresponding Focal Plane Module A (FPMA) and B (FPMB) on May 10, 2019  for a total of $\sim119.2$ ks of elapsed time.  The Level 1 data products were processed with the {\it NuSTAR} Data Analysis Software (NuSTARDAS) package (v. 1.8.0). Cleaned event files (level 2 data products) were produced and calibrated using standard filtering criteria with the \textsc{nupipeline} task and the latest calibration files available in the {\it NuSTAR} calibration database (CALDB 20190410). Extraction radii for the source and background spectra were $60$ arcsec and $70$ arcsec and the net exposure times for the two time-averaged spectra are and 57.5 and 57.1 ks for the FPMA and FPMB, respectively. The two {\it NuSTAR} spectra were binned in order to over-sample the instrumental resolution by at least a factor of 2.5 and to have a SNR greater than 3$\sigma$ in each spectral channel, for the spectral fitting procedure. We then adopted a linear time sampling of 5.8 ks, for a total number of 20 pairs of spectra. The cross-calibration factors between the two detectors were found to be within $2\%$ in each time slice. A further constant was added (within $10\%$ throughout the observation), to consider XMM-{\it NuSTAR} cross-calibration uncertainties.

\begin{figure*}
 \epsfig{file=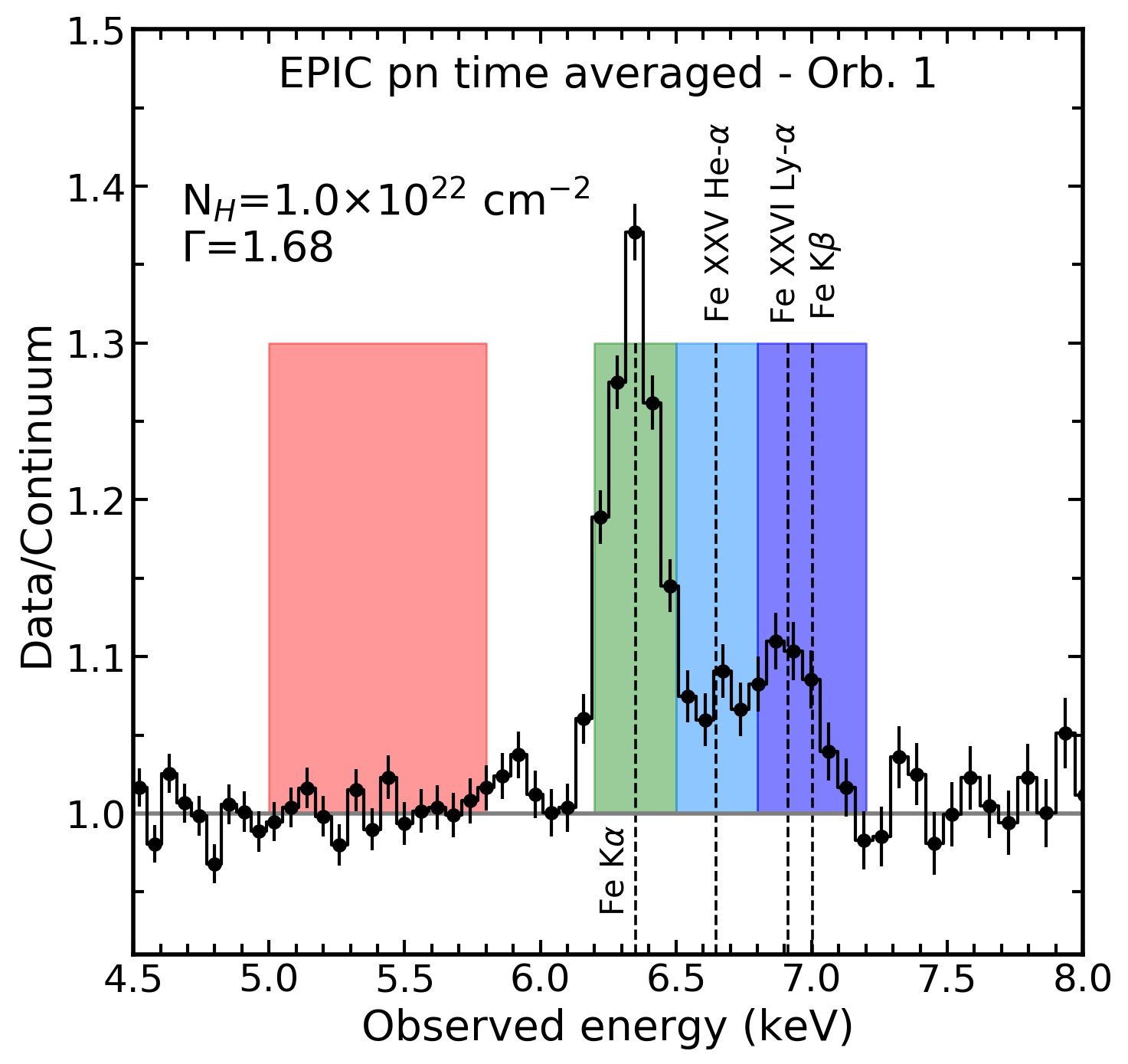, width=0.68\columnwidth}
 \epsfig{file=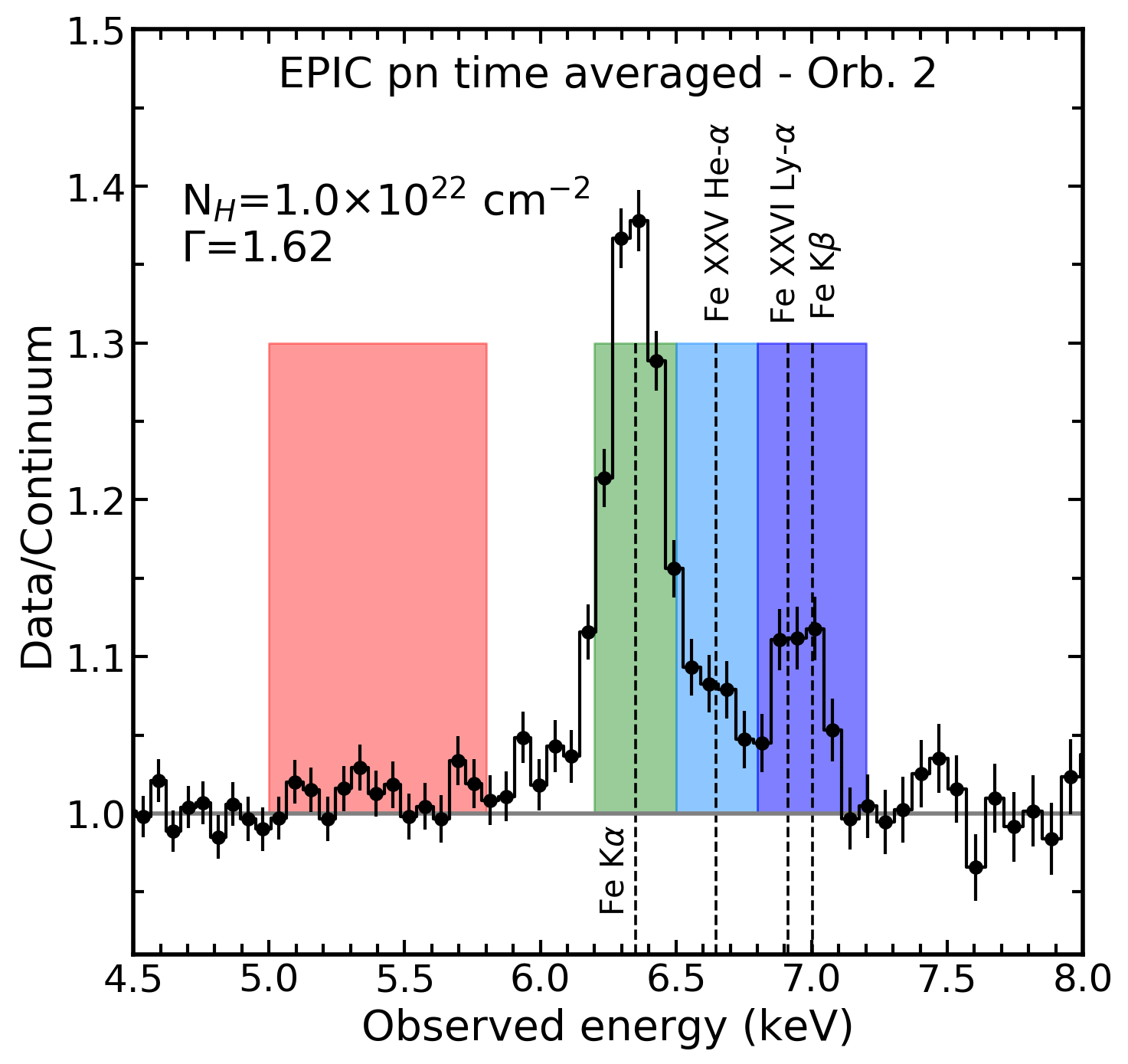, width=0.68\columnwidth}
 \epsfig{file=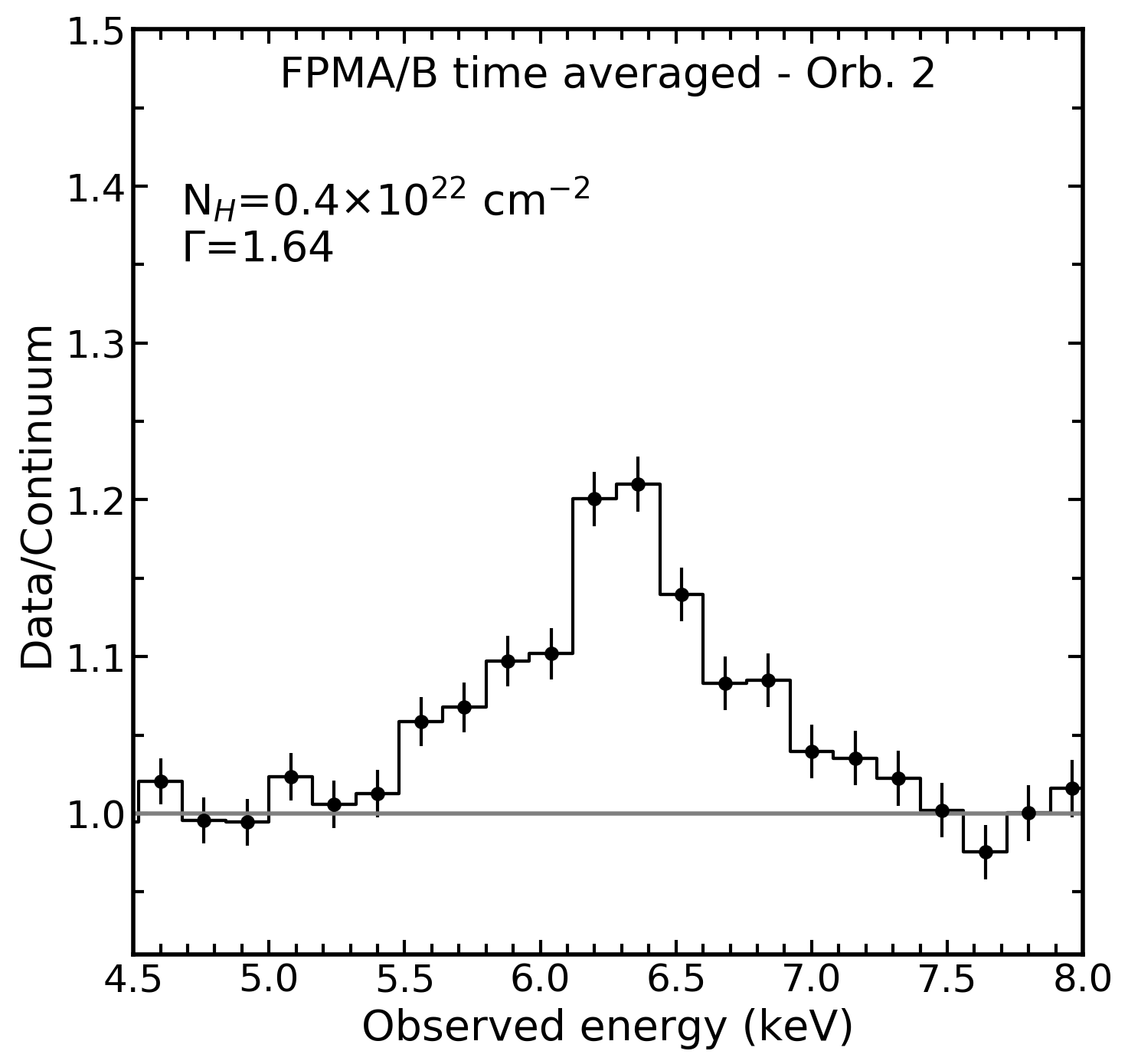, width=0.68\columnwidth}
  \caption{The ratios between the time-averaged data and the associated best fitting continuum models are shown, for the EPIC pn and FPMA/B observations. The continuum is composed of an absorbed power law fitted between 3-5 keV plus 8-10 keV, the best fitting values for the column density and photon index are reported in the top-left corner. The shaded regions indicate the four energy bands used for the variability patterns and the dashed lines show the emission lines between  6-7 keV included in the phenomenological spectral analysis. }
  \label{t_avg}
\end{figure*}

\section{Data analysis}
\subsection{Estimates of the black hole mass}
The EPIC pn light curves, extracted from the 2-10 keV energy band, are plotted in Fig. 1 and they show a count rate which ranges from 7.78$\pm$0.11 to 12.33$\pm$0.14 cts/s. They can be used to calculate a normalized excess variance $\sigma^2_{\rm rms}=1.9\pm1.0\times10^{-3}$ \citep{nan97}, adopting time bins of 250 s and selecting segments of 20 ks. Assuming the $\sigma^2_{\rm rms}-{\rm M_{bh}}$ correlation reported in \citet{ppb12} we estimate a black hole mass M$_{\rm BH}$=(3.0${}^{+5.5}_{-1.5})\times10^7$ M$_{\odot}$, including also intrinsic uncertainties on the relation itself. The value obtained from the M$_{\rm BH}-\sigma_{\star}$ relation is M$_{\rm BH}$=4.8${}^{+3.9}_{-2.4}\times10^7$ M$_{\odot}$), using a bulge stellar velocity dispersion $\sigma_{\star}=158\pm13$ km/s \citep{nw95} and the relation from \citet{grg09}. Given the good agreement between the M$_{\rm BH}$ inferred via the excess variance technique and the one derived from the M$_{\rm BH}-\sigma_{\star}$ relation, we will use a value of M$_{\rm BH}$=$3.0^{+5.5}_{-1.5}\times10^7$ M$_{\odot}$ for the black hole mass throughout the paper.

\subsection{Time variability and excess maps}
In the following, we will consider data in the 2-10 keV and 3-80 keV energy ranges for XMM and {\it NuSTAR} spectra. The detailed analysis of the broadband data set will be the subject of a different paper (Marinucci et al., in prep.).
We first fitted the time-averaged spectra with a model composed of an absorbed power law ({\sc zwabs}$\times${\sc pow} in {\sc Xspec}) multiplied by a Galactic absorption component ({\sc TBabs}) with N$_{\rm H}=4.8 \times 10^{20}$ cm$^{-2}$ \citep{kalberla05}, excluding the energy range dominated by the Fe K lines (5-8 keV). The ratios between the time-averaged data and the best fitting continuum models are plotted in Fig. \ref{t_avg}, once the 5-8 keV band is included. A cross-calibration constant is added to the FPMA/B spectral fit and they are then simultaneously plotted using the {\sc setplot group} command within {\sc Xspec}. The energy binning of the spectra is described in Sect. 2 and it is not the one used for creating the excess maps. Residuals are present on both the red and blue sides of the narrow iron K$\alpha$ emission line, in all the three data sets. The main spectral changes are in the 5-6 keV and 6.5-7.2 keV bands, with clear variations between the first and the second XMM orbit.
\begin{figure*}
 \epsfig{file=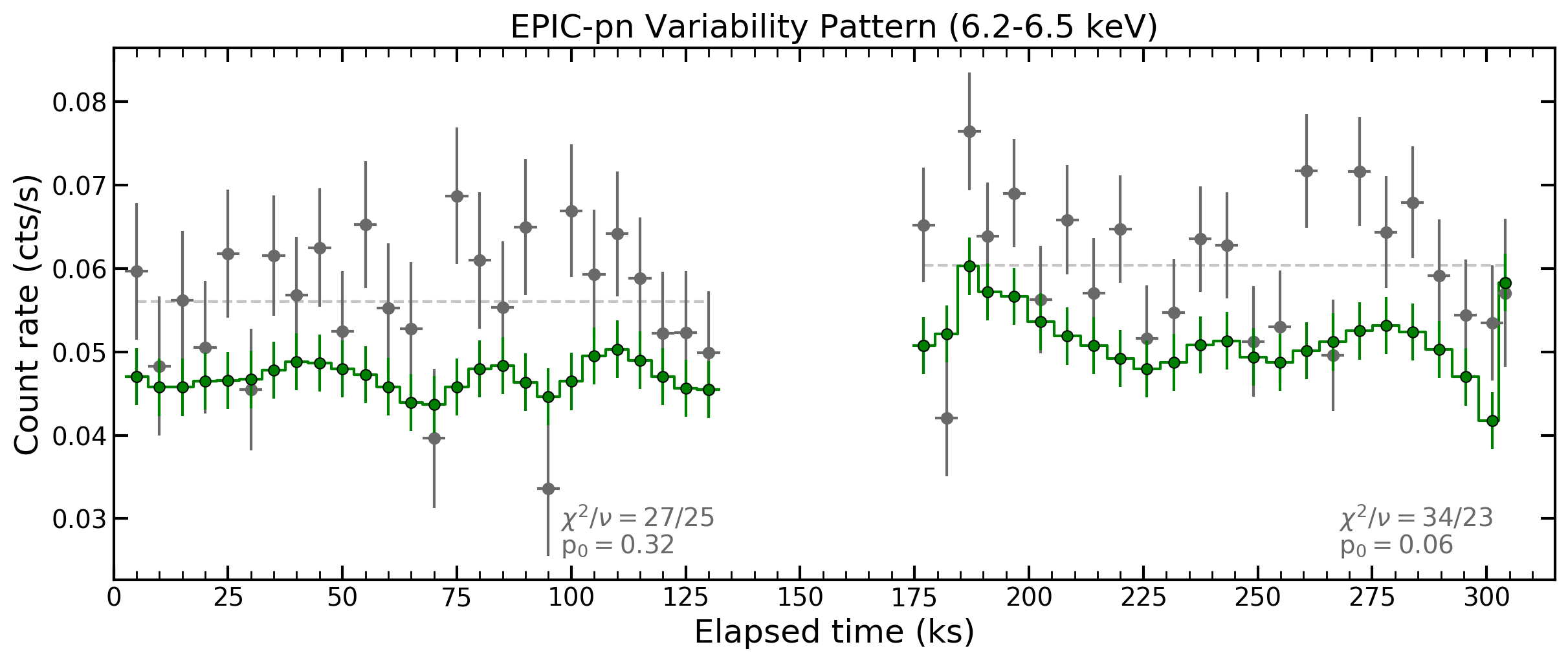, width=1.04\columnwidth}
 \epsfig{file=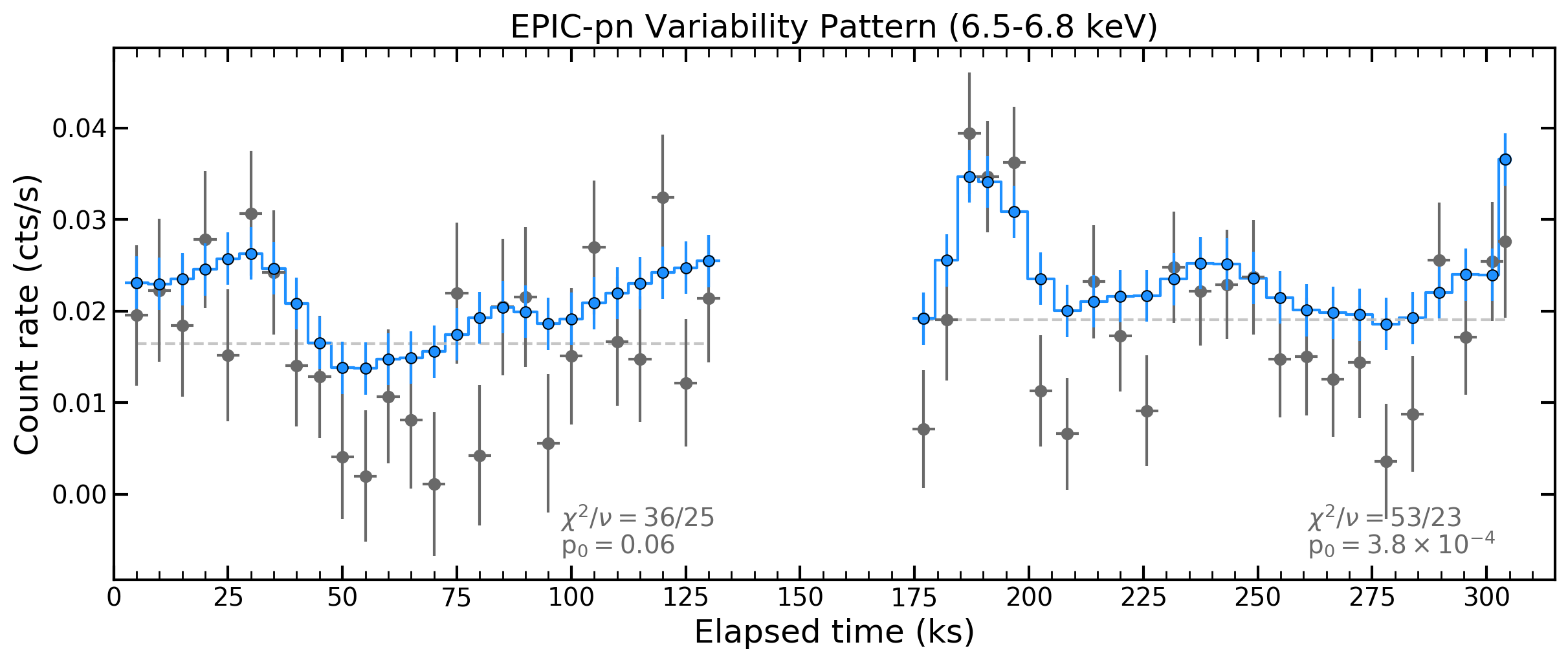, width=1.04\columnwidth}
  \epsfig{file=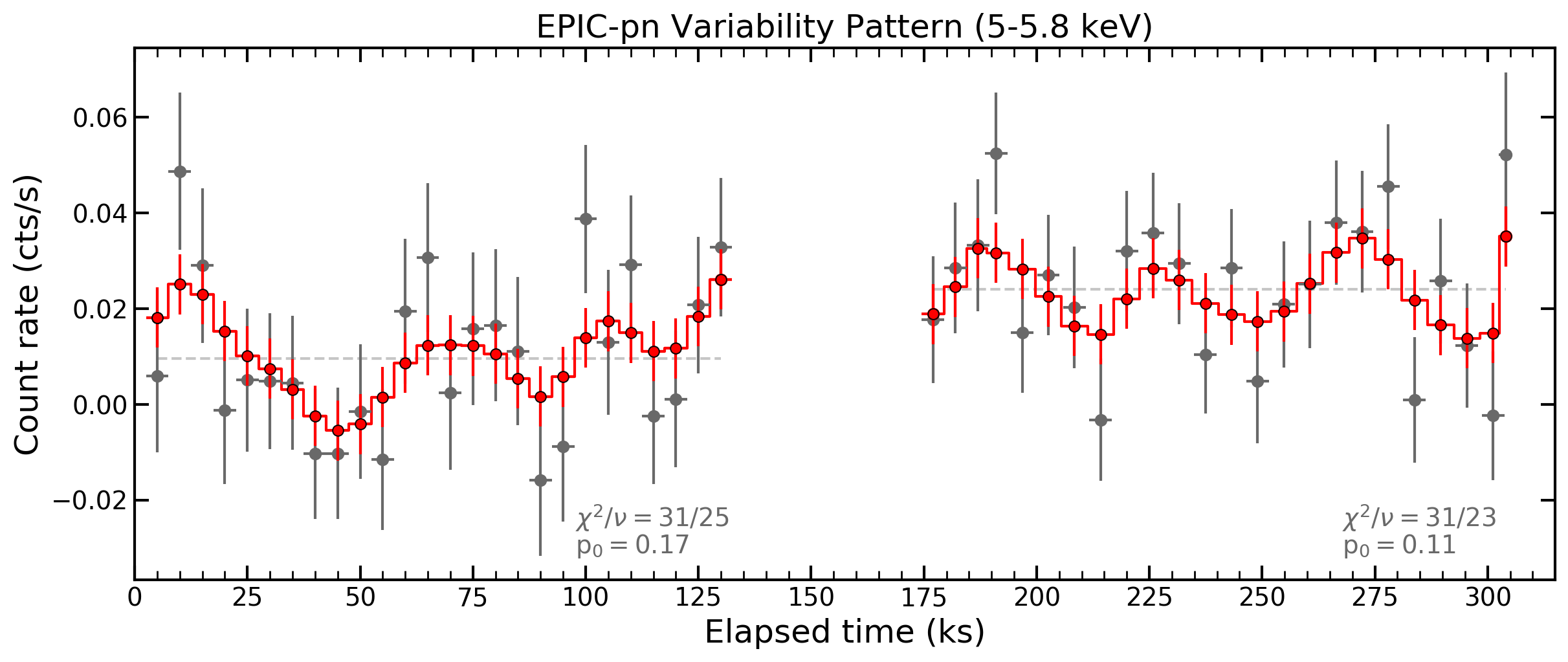, width=1.04\columnwidth}
 \epsfig{file=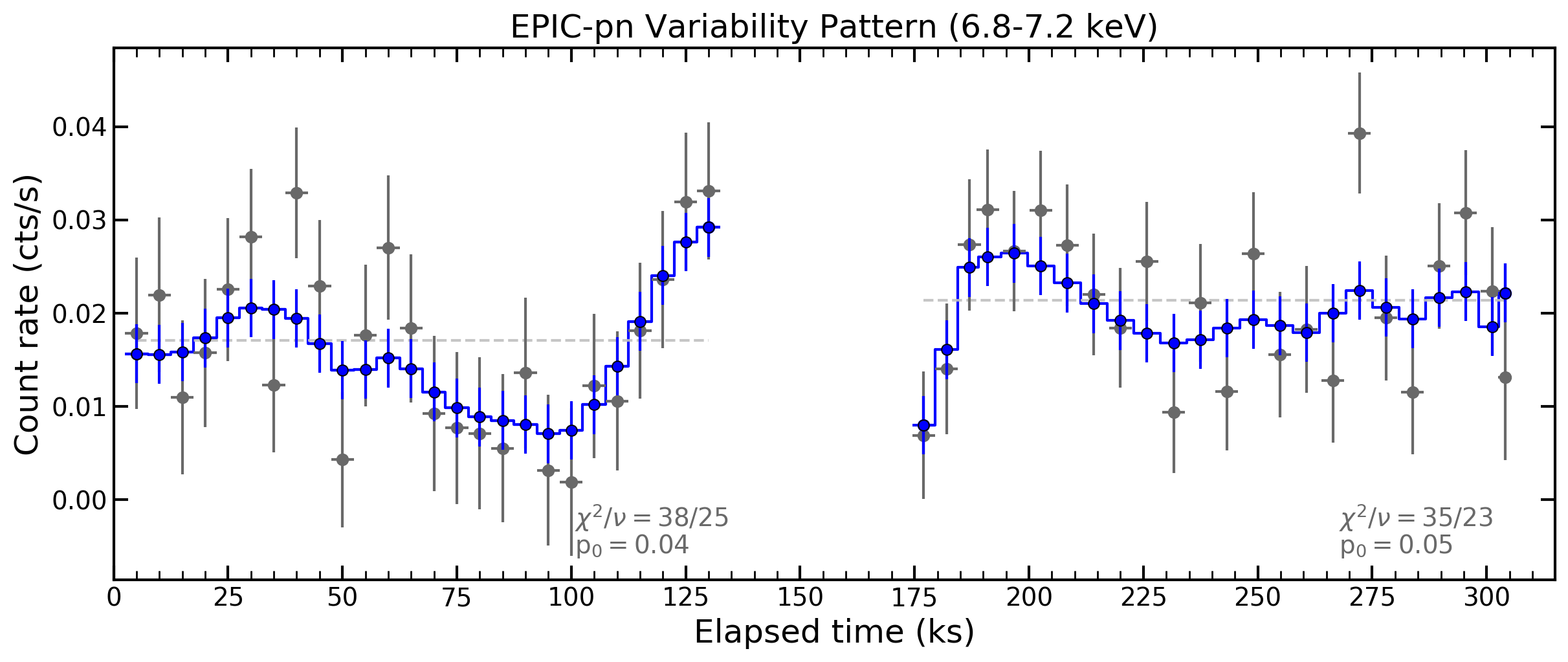, width=1.04\columnwidth}
 \caption{The smoothed excess count rates in the 6.2-6.5 keV, 6.5-6.8 keV, 5.0-5.8 keV, 6.8-7.2 keV bands are shown in green, light blue, red and dark blue, respectively. Unsmoothed data are shown in grey. The best fitting constant functions are shown as dashed grey lines.}
  \label{variability}
\end{figure*}
 \begin{figure*}
 \epsfig{file=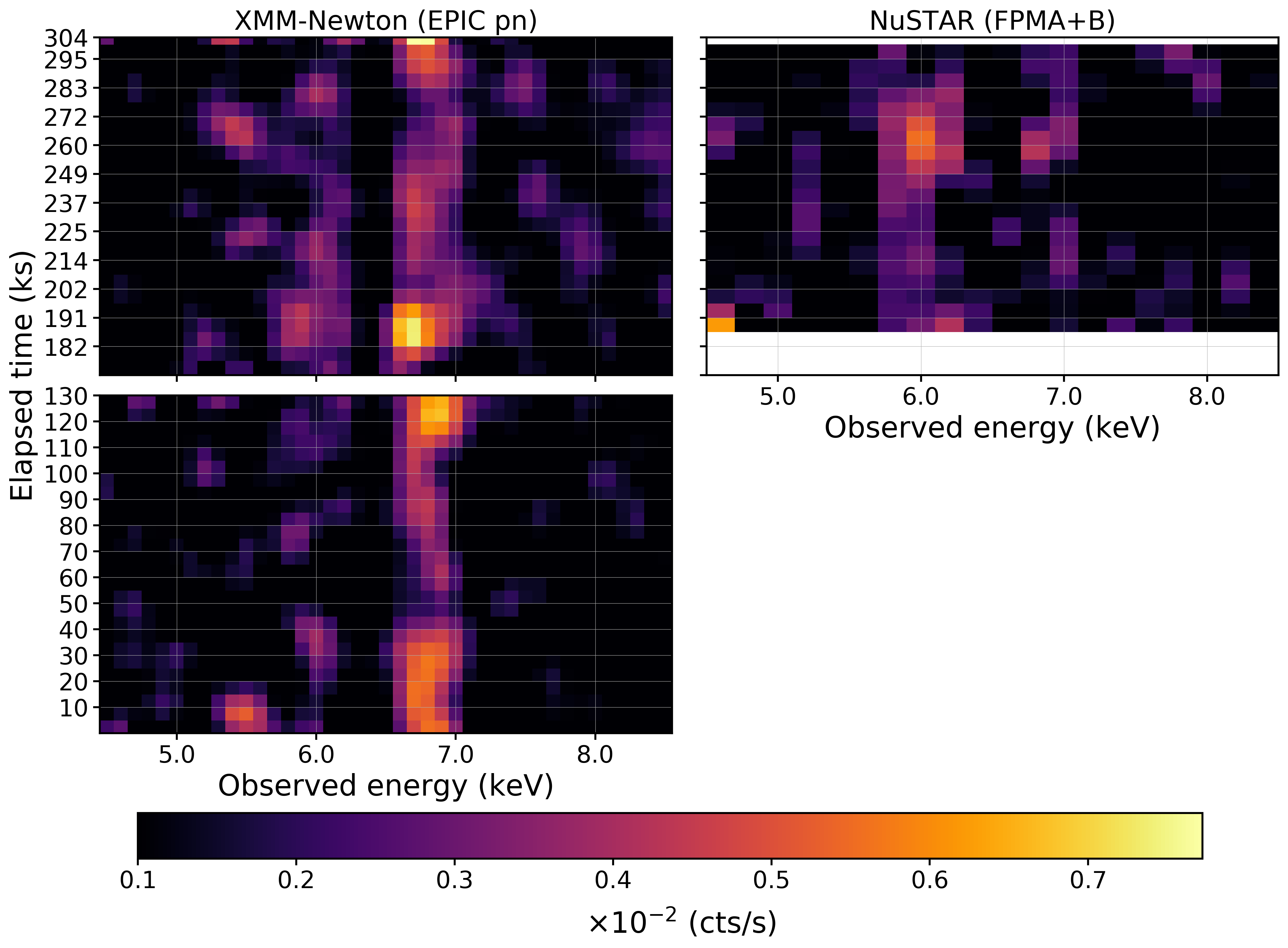, width=1.9\columnwidth}
 \caption{Excess emission map (in units of cts/s) for the full observation, when a model composed of a variable continuum and narrow iron K$\alpha$/K$\beta$ emission lines at 6.4 and 7.058 keV are considered.}
  \label{xmm_noKa}
\end{figure*}
 
We then applied this model for the continuum to each of the 50 spectral slices.
XMM and {\it NuSTAR} spectra are fitted simultaneously when simultaneous data are available (time intervals between 191 ks and 301 ks), with the addition of a multiplicative constant to account for cross-calibration uncertainties. No residuals in excess to the best fit models are observed above 10 keV, suggesting a lack of a significant Compton hump due to cold reflection. Best fit models are stored and data with constant energy bins (100 eV for XMM and 200 eV for {\it NuSTAR}) are loaded. For each spectrum we calculate the the excess of counts with respect to the corresponding model and plot it as a function of energy and time.  Any appreciable intensity variations are expected to occur on longer timescales than the sampling time (5-5.8 ks), therefore a smoothing procedure is expected to suppress the noise between adjacent pixels, providing a cleaner view of the possible variability patterns. Following the method described in \citet{imf04} and \citet{npr16} we smoothed the excess emission map through an elliptical Gaussian kernel with ($\sigma_{\rm E}$, $\sigma_{\rm t}$)=(1.06, 1.27) pixels for the first XMM orbit and ($\sigma_{\rm E}$, $\sigma_{\rm t}$)=(1.06, 1.10) for the second one: corresponding to a FWHM of 250 eV$\times$15 ks. The different width in time $\sigma_{\rm t}$ is due to the larger time intervals (5.8 ks rather than 5 ks) adopted in the second XMM orbit. FPMA/B detectors have a worse energy resolution (400 eV at 6 keV) than the EPIC pn and we used ($\sigma_{\rm E}$, $\sigma_{\rm t}$)=(0.531, 1.10) pixels. For an energy binning of 200 eV and an integration time of 5.8 ks these values correspond to a FWHM of 250 eV$\times$15 ks. We plotted, in the four panels of Fig. \ref{variability}, the energy-integrated count rates in excess to the continuum.  We note that the energy bands of the smoothed data, due to the 250 eV width of the smoothing function, are wider than the ones used for the unsmoothed data. Counts integrated between 5 and 5.8 keV were associated to the {\it red flare}, between 6.2-6.5 keV to the narrow core of the iron K$\alpha$, between 6.5-6.8 keV to the {\it blue flare I} and between 6.8-7.2 keV to the {\it blue flare II}. The different energy bands, over-imposed to the time averaged spectra, can be seen in Fig. \ref{t_avg}. Grey data points are used for plotting the excess count rates before the smoothing process in Fig. \ref{variability}. We note that the smoothing process underestimates the 6.2-6.5 keV counts and overestimates the 6.5-6.8 keV counts (top panels in Fig. \ref{variability}): this is an effect due to the 250 eV energy smoothing applied to the adjacent energy pixels. Assuming that the observed counts follow a Poissonian distribution ($\sigma_C=\sqrt{C}$), we calculated their error bars as the root sum of the squares of the errors extrapolated from the best fits of the continuum and the ones on the counts in excess. For comparison, we find $3282\pm57$ counts for the continuum and $171\pm13$ counts in excess in the 5-5.8 keV band, in the 10 ks spectral slice. During the second orbit, at 278 ks, we obtain $2803\pm53$ counts and $184\pm14$ counts for the continuum and for the excess, respectively. When we apply the smoothing filter, we loose the counting statistics and we therefore run extensive simulations to estimate the error on the plotted ratios, as detailed below. Following the procedures presented in \citet{imf04} and \citet{dmi09}, we simulated N$_{\rm sim}$=1000 time-energy maps with constant components in the four energy bands (obtained from fitting four Gaussians to the time-averaged spectra) with the associated best fit continuum model. Since the flux of the four spectral components ({\it red flare}, narrow iron K$\alpha$, {\it blue flare I} and {\it blue flare II}) is fixed in each simulation, the variance of the individual light curves after the smoothing serves as the measurement uncertainty. We therefore considered the mean variance of the 1000 light curves and the standard deviation was assumed as the measurement error.
 To estimate the variability of the excess counts in the four energy bands we applied a constant model to the unsmoothed data points and the corresponding $\chi^2/\nu$ values and null-hypothesis probabilities $p_0$ are reported in Fig. \ref{variability}. The largest deviation from a constant model is observed for the {\it blue flare I} component, during the second orbit.
 
At last, we added two narrow Gaussian components to reproduce the constant iron K{$\alpha$}/K$\beta$ emission lines (with free energy and normalization for the former) to the XMM+{\it NuSTAR} continuum model and follow the previous steps to reproduce a count rate excess map. We will confirm in the next Section that this spectral components are statistically consistent with being constant throughout the observation. The result is shown in Fig. \ref{xmm_noKa}. This is the first excess emission map constructed using {\it NuSTAR} data and, despite the lower spectral resolution and the non consecutive on-source spectra, a flux modulation in the 5.5-6.5 keV band can still be observed. However, for these two reasons, we will only consider XMM data hereinafter.

\subsection{$F_{\rm var}$ spectrum}
\begin{figure}
 \epsfig{file=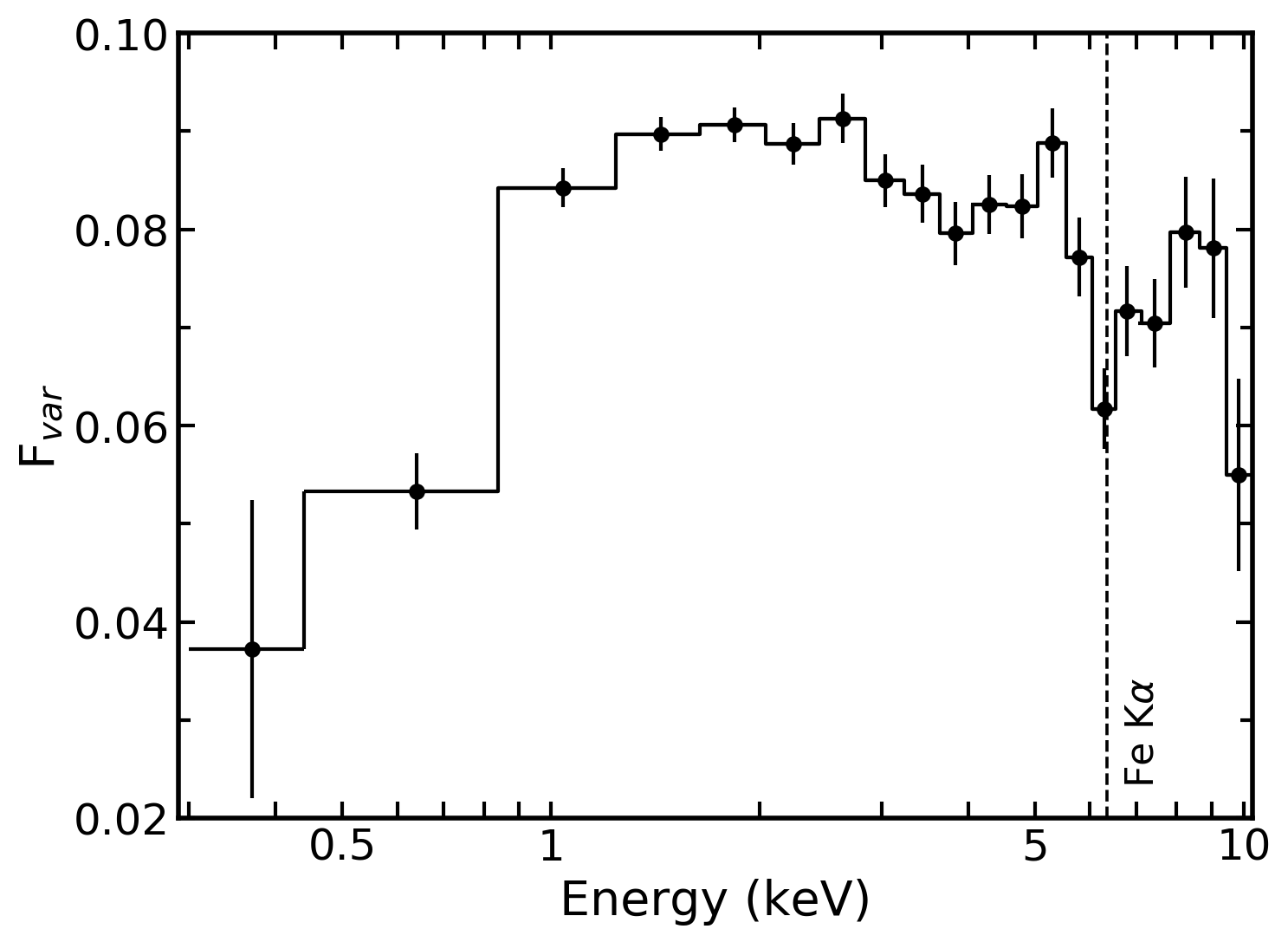, width=\columnwidth}
 \caption{$F_{\rm var}$ spectrum of the full EPIC-pn observation. The dashed line indicates the neutral Fe K$\alpha$ observed energy.}
  \label{Fvar}
\end{figure}

Following \citet{vaugh03}, we computed the fractional root mean square (rms) variability amplitude \citep[$F_{\rm var}$:][]{ponti04,ponti06}. To this aim we extracted light curves of each observation, in small energy bins (with a time bin of 50 s) and computed the Poisson-noise subtracted power spectrum \citep[normalised to units of squared fractional rms,][]{mkk91}. We integrated each power spectrum over the frequency range $0.9\times10^5-10^4$ Hz, corresponding to time scales ranging between 110 ks (approximately covering the entire duration of a single observation) and 10 ks. These time scales are chosen so as to sample the observed modulations in the Fe K line complex. From the square root of the integrated power we derived an estimate of the $F_{\rm var}$ as a function of energy, for each observation. Estimates from the two observations were then averaged to obtain the $F_{\rm var}$ spectrum displayed in Fig. \ref{Fvar}. The $F_{\rm var}$ spectrum shows a spectral shape typical of absorbed sources \citep[e.g.][]{dap20}, with the soft energy drop due to the presence of constant spectral components which dominate this part of the spectrum. At higher energies ($>1$ keV) the $F_{\rm var}$ increases ($\sim$7-9\%), with a peak at $\sim$2 keV and slightly decreasing towards harder energies. The sharp drop of $F_{\rm var}$ at E$\sim$6.4 keV (marked by the vertical dashed line) is due to the presence of a constant neutral Fe K$\alpha$ emission line. At slightly lower energies (E$\sim$5-6 keV) the $F_{\rm var}$ shows instead an increase, which hints at presence of enhanced variability associated with redshifted Fe K line components. This result is in agreement with the behavior observed in the excess map.

\begin{figure}
 \epsfig{file=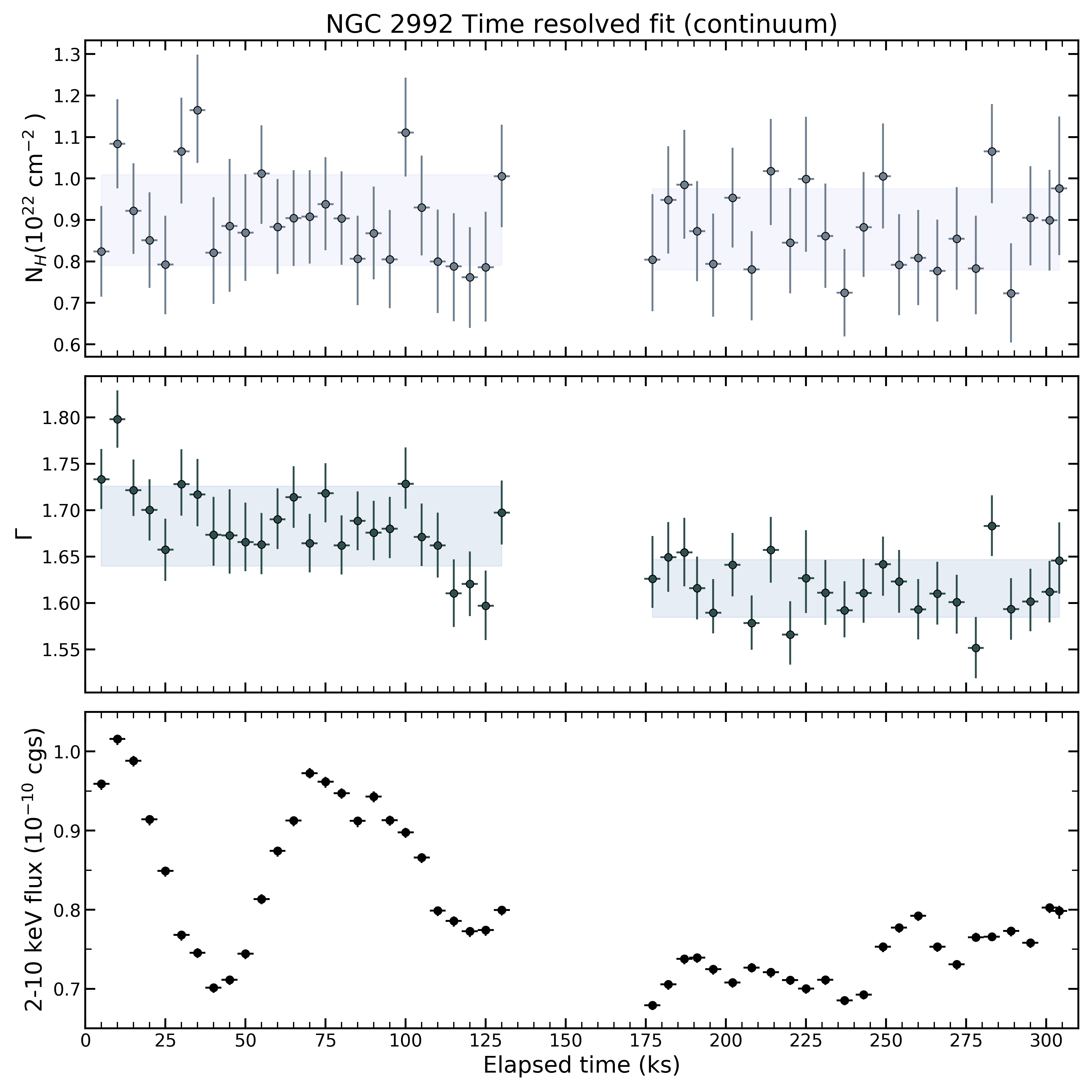, width=\columnwidth}
 \epsfig{file=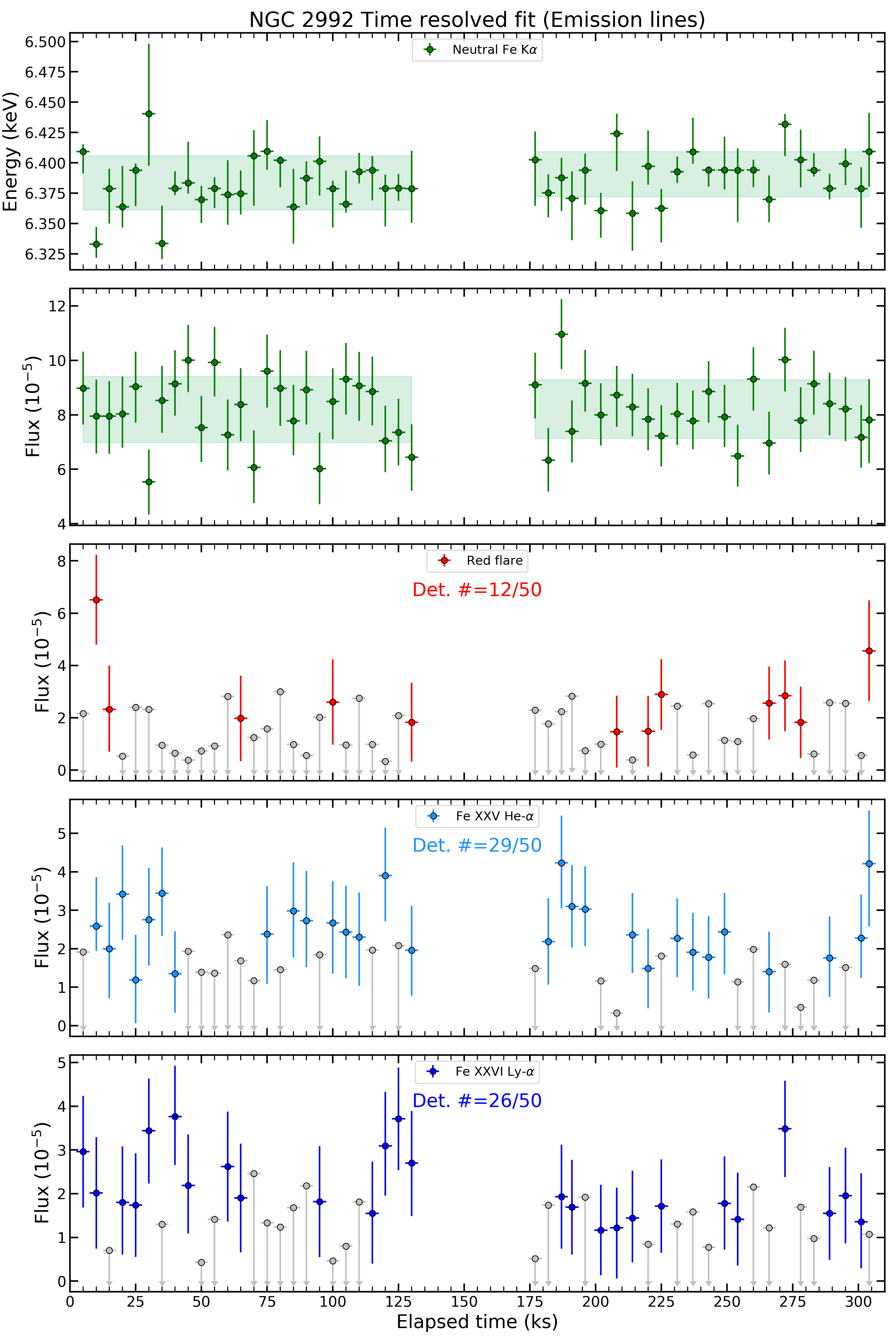, width=\columnwidth}
 \caption{ Best fit values for the continuum (top-panels) and emission lines (bottom-panels) when the phenomenological model described in Sect. 3.3.1 is applied to the XMM data set. Fluxes are in ph cm$^{-2}$ s$^{-1}$ units and shaded regions indicate 1$\sigma$ above and below mean values. Errors are calculated using 68\% c.l. The number of intervals in which the emission lines are detected is reported in the top center of bottom panels.}
  \label{fit}
\end{figure}
\subsection{Spectral fitting}
In the following, we describe the 2-10 keV spectral fitting of the 50 EPIC pn spectra. Our aim is to characterize and explain the observed variability patterns around the Fe K complex, using both a phenomenological and a self-consistent physical model. We first model the counts excesses in the four energy bands shown in Fig. \ref{variability} with variable Gaussians and then with flaring spots from the accretion disk.

\subsubsection{Phenomenological analysis}
The phenomenological model applied to the data set is composed of the absorbed power law considered in Sect. 3.1 and five narrow Gaussian lines, to reproduce the neutral Fe K$\alpha$ and K$\beta$, , the Fe {\sc xxv} He-$\alpha$ and the Fe {\sc xxvi} Ly-$\alpha$ emission lines at 6.4 keV, 7.058 keV, 6.7 keV and 6.966 keV, respectively. One additional Gaussian is included in the model to reproduce the {\it red flare}.  The normalization of the Fe K$\beta$ line is fixed to 0.16$\times$N$_{\rm K\alpha}$ \citep{mbm03}. The free parameters in the fits are the column density  N$_{\rm H}$, the photon index $\Gamma$, the energy centroid of the neutral Fe K$\alpha$ and the normalization of the power law and of the four emission lines. The energy centroid of the spectral component associated to the {\it red flare} is fixed to 5.4 keV. We show the best fitting values in Fig. \ref{fit}, errors are calculated using a 68 \% confidence level.  The shaded regions indicate 1$\sigma$ above and below the mean N$_{\rm H}$, $\Gamma$, energy centroid and flux of the neutral iron K$\alpha$. The number of detected emission lines is reported in the top-center of the three bottom panels. The minimum and maximum flux levels measured throughout the observation are F$_{2-10}=(6.8\pm0.1)\times10^{-11}$ erg cm$^{-2}$ s$^{-1}$ and F$_{2-10}=(1.01\pm0.01)\times10^{-10}$ erg cm$^{-2}$ s$^{-1}$, corresponding to luminosities L$_{2-10}=9.5\times10^{42}$ erg s$^{-1}$ and L$_{2-10}=1.5\times10^{43}$ erg s$^{-1}$ (corrected for intrinsic absorption), respectively. Adopting the bolometric correction $K_X(L_X)$ from \citet{dbr20} and a black hole mass M$_{\rm BH}$=$3\times10^7$ M$_{\odot}$, we obtain an accretion rate interval L$_{\rm Bol}$/L$_{\rm Edd}\simeq$4-6\%.

\begin{figure}
  \epsfig{file=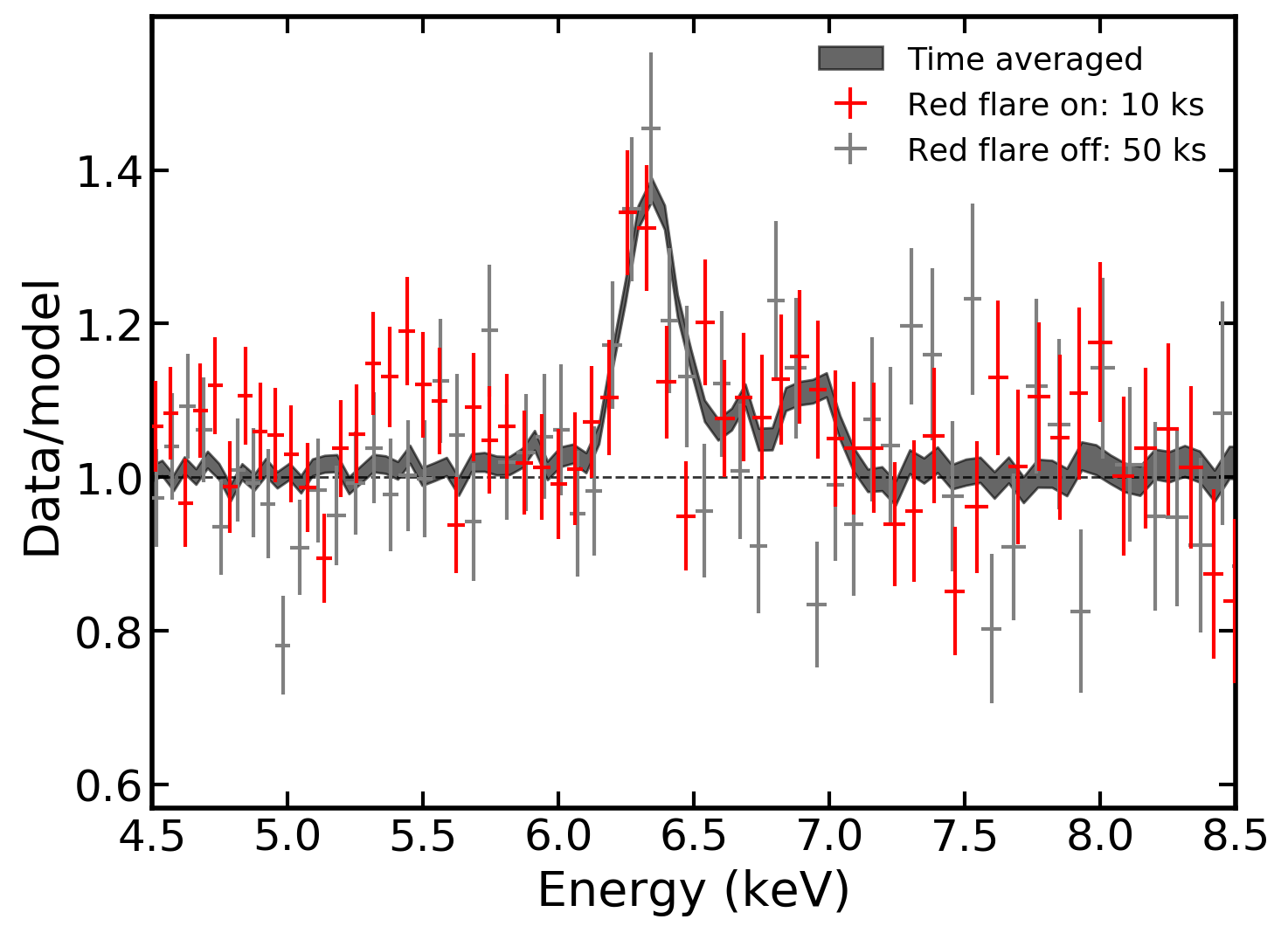, width=\columnwidth}
\epsfig{file=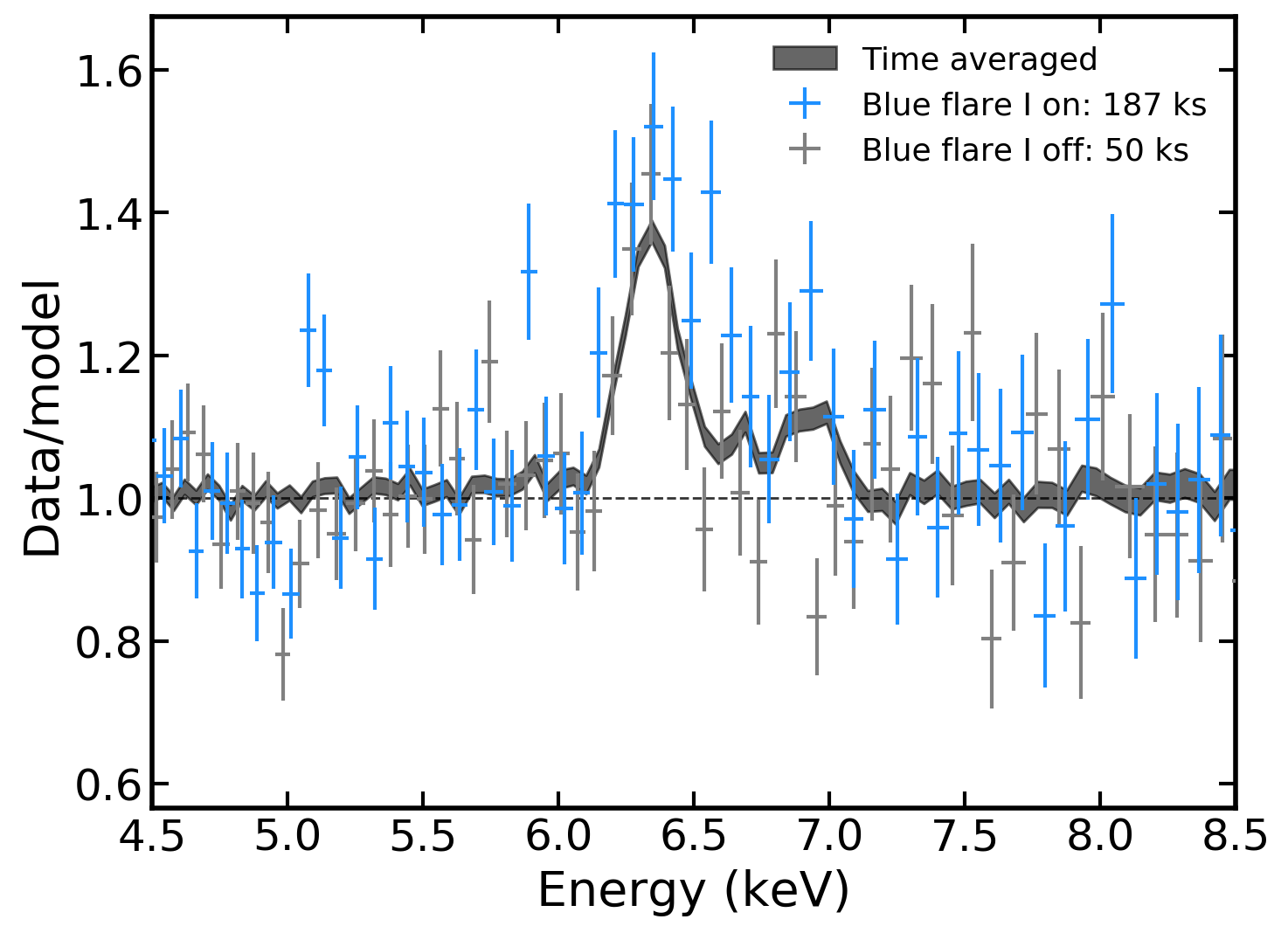, width=\columnwidth}
 \caption{The ratios between the data and the absorbed power law model is shown for the time averaged spectrum (black shaded region) and for the two spectral slices corresponding to 10 ks (red data points, top panel), 187 ks (light blue data points, bottom panel) and 50 ks (grey data points), which are representative of the two flaring and one quiescent state. }
  \label{ratio_All}
\end{figure}

While the neutral Fe K$\alpha$ normalization is statistically consistent with being constant throughout the observation,  variations of the other emission lines are detected. The maxima of the {\it red flare} normalization occur at 10 ks and 304 ks. For the sake of a visual comparison, we plot in the top panel of Fig. \ref{ratio_All} the ratio between the data and the absorbed power law model for the time averaged spectrum (185.5 ks long) and for the two spectral slices corresponding to 10 ks and 50 ks, in which the maximum and an upper limit are retrieved for the {\it red flare} component. We show in the bottom panel of Fig. \ref{ratio_All} the same ratios but for the maximum of the {\it blue flare I} component. The maxima of the Fe {\sc xxv} He-$\alpha$ and Fe {\sc xxvi} Ly-$\alpha$ components occur at 187 ks and 40 ks, respectively. These values are in perfect agreement with our results from the excess map technique (Fig. \ref{variability} and Fig. \ref{xmm_noKa}). The timescales of the detected variations of the four phenomenological emission lines are consistent with distances of $\sim30$-$40$ r$_g$ (for light-crossing times of 5-6 ks). This motivates the usage of a self-consistent model in which the transient Fe K lines are due to various orbiting flaring regions above the accretion disk.

\subsubsection{The {\sc KYNrline} model}
In this Section, we apply the {\sc KYN} model \citep{dov04} to the 50 EPIC pn spectra. The change of the emission lines amplitude and energy is explained in terms of orbital motion in a relativistic gravitational field close to the central black hole. The model assumes a space-time around the black hole which is described by the Kerr metric and a Keplerian, geometrically thin and optically thick accretion disk\footnote{a full description of the model can be found at \url{https://projects.asu.cas.cz/stronggravity/kyn/}}. In particular, we use the {\sc KYNrline} model component in {\sc Xspec}, which reproduces a relativistic line with a broken power-law radial emissivity. The emitting regions are non-axisymmetric, i.e. only part of the disc may be emitting (sections in radius and azimuth). 
\begin{figure}
  \epsfig{file=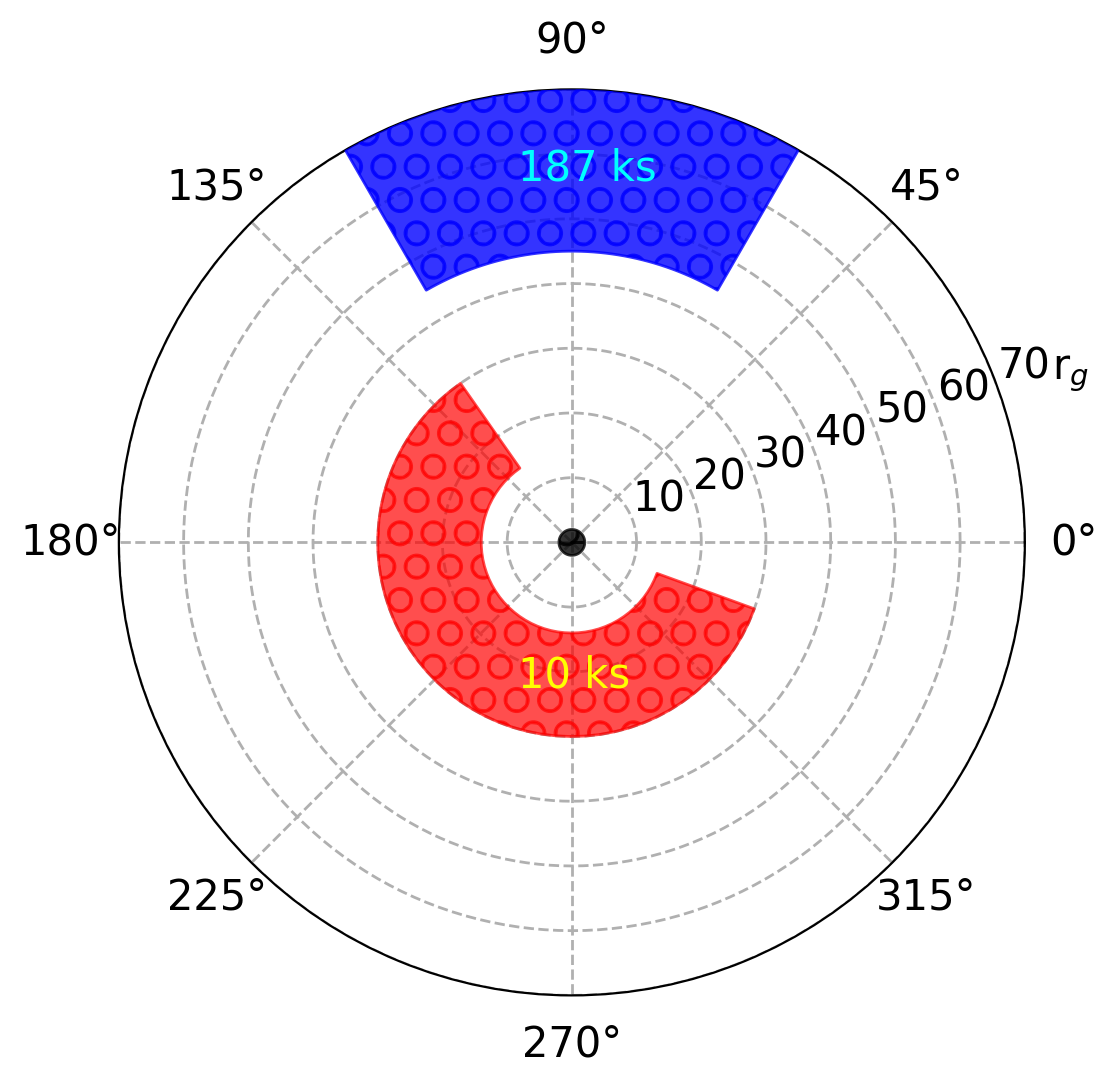, width=\columnwidth}
 \caption{Schematic view of the emitting regions associated to the two {\sc KYNrline} components described in Sect. 3.3.2. The red area indicates the best fitting values which reproduce the {\it red flare} and the {\it blue flare  II } at 10 ks, when the 5-5.8 keV excess is maximum. The blue one shows the best fitting values for the {\it blue flare I} at its maximum, during the 187 ks time interval.}
  \label{accdisk}
\end{figure}

\begin{figure*}
 \epsfig{file=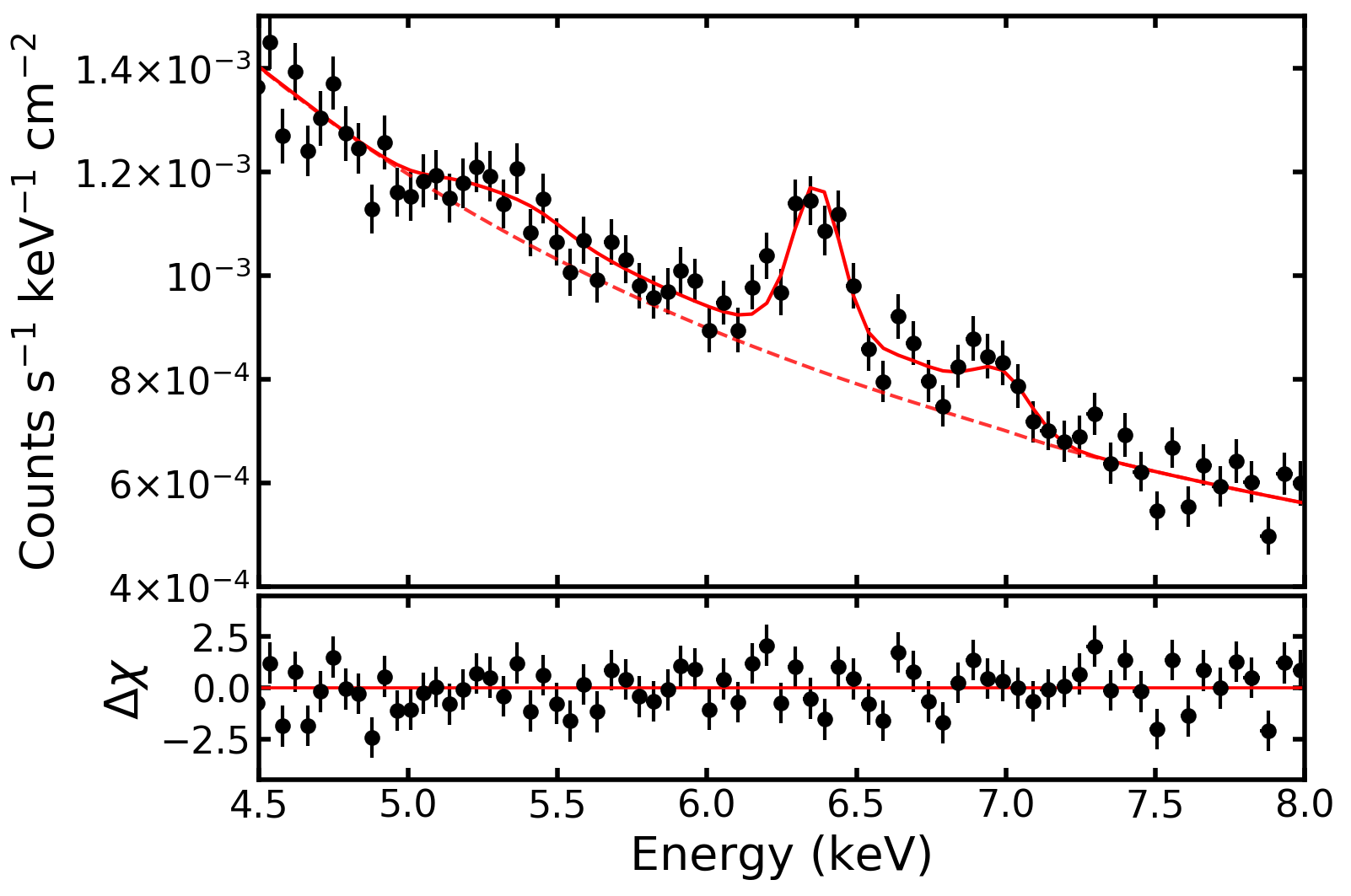, width=\columnwidth}
  \epsfig{file=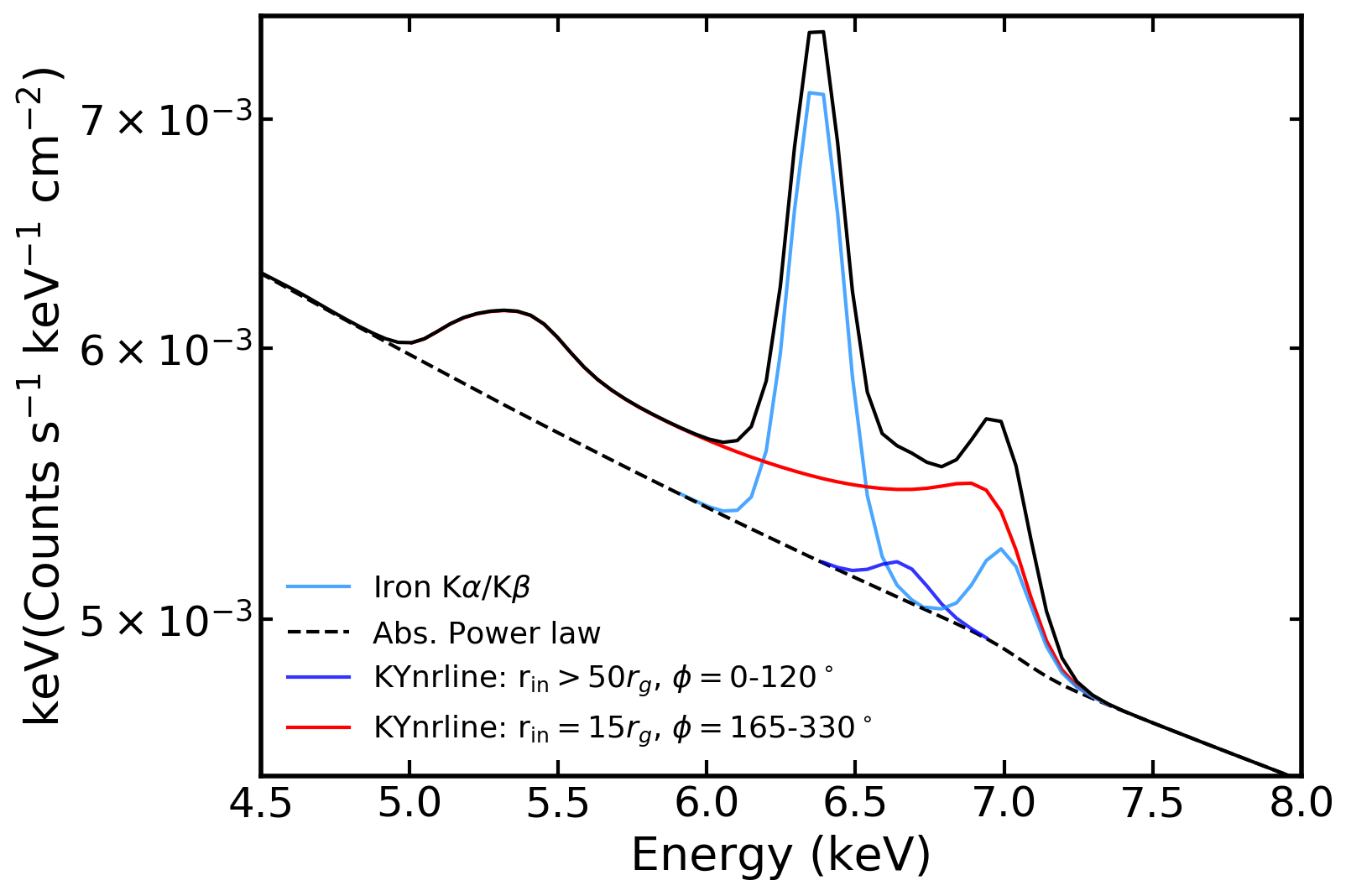, width=\columnwidth}
 \caption{{\it Left panel:} the {\it flaring mode} spectrum and the best fitting model are plotted with the corresponding residuals. The spectrum is the sum of the ones extracted from time intervals peaking at 191, 225, 272 and 304 ks. {\it Right panel:} the final {\sc KYNrline} model applied to the 50 EPIC pn spectra is shown in black, in the 4.5-8 keV band only. This represents the best fitting model of the {\it flaring mode} spectrum only, the different components are free  in each spectral slice. The dashed black line indicates the absorbed power law component and the narrow iron K$\alpha$ and K$\beta$ emission lines are plotted in cyan. We show the {\it red flare} + {\it blue flare II} ($r_{\rm in}=15\pm3$ r$_g$, $\phi=165^{\circ}-330^{\circ}$) and {\it blue flare I} ($r_{\rm in}>50$ r$_g$, $\phi=0^{\circ}-120^{\circ}$) components as red and blue solid lines, respectively.}
  \label{redadded}
\end{figure*}

We first adopt a model which is composed of the absorbed power law considered in Sect. 3.1 and two narrow Gaussian lines, to reproduce the neutral Fe K$\alpha$ and K$\beta$. The parameters $\Gamma$, N$_{\rm H}$ and normalization of the continuum are free. The Fe K$\beta$ is always included in the following fits and its normalization is fixed to 0.16$\times$N$_{\rm K\alpha}$. The neutral Fe K$\alpha$ energy centroid and normalization are free. We then included a {\sc KYNrline} component, fixing the emissivity of the disk to $\epsilon(r)\propto r^{-3}$, the black hole spin to $a^*=0.998$ and the annular region extension to 10 r$_g$. In our fits, we chose a framework where positive angular velocity corresponds to counterclockwise rotation and $\phi=0^{\circ}, 360^{\circ}$ to the maximal Doppler blueshift for matter moving toward the observer \citep[same convention as in][]{dov04, npr16}. We also assumed an energy at rest E$_{\rm R}=6.7$ keV and an inclination angle of the accretion disk $i=40^{\circ}$ with respect to our line of sight \citep{ymg07}.  The properties of the emitting annulii are therefore estimated by leaving the inner radius r$_{\rm in}$, the angle $\phi$ and the angular extension $\Delta\phi$ free in the fits. For simplicity, the reported uncertainties on the angular sizes are the lower errors on $\phi$ and the upper errors on $\Delta\phi$. This first {\sc KYNrline} spectral component reproduces the 5.4 keV and the 7.0 keV observed peaks ({\it red flare} + {\it blue flare II}). A second {\sc KYNrline} component is then included in the model to account for the 6.7 keV excess ({\it blue flare I}).

This model is applied to the 50 EPIC pn spectral slices. Best fit values for the primary continuum, neutral Fe K$\alpha$ line, annular extension, size and normalization of the {\sc KYNrline} components are reported in Table 1. All the spectra, the corresponding best fit models and the relative residuals, are shown in Fig. \ref{bfit1} and \ref{bfit2}. We plot in Fig. \ref{accdisk} a sketch of the emitting regions associated to the two {\sc KYNrline} components for the 10 ks and 187 ks time intervals. 

\section{Discussion}
The EPIC-pn time variability patterns presented in Sect. 3.2 have shown several transient emission features throughout the 5-7 keV band and some recursive flares in the 5-5.8 keV band can be seen in Fig. \ref{variability} and \ref{xmm_noKa}. If a sinusoidal function is applied to the unsmoothed data of the second XMM orbit (Fig. \ref{variability}, bottom-left panel) a best fitting orbital period $T=41\pm1$ ks is retrieved, with a corresponding $\chi^2/\nu$=20/20. Assuming a maximally rotating black hole spin and a black hole mass M$_{\rm BH}$=$3.0^{+5.5}_{-1.5}\times10^7$ M$_{\odot}$, we can use the relation from \citet{bpt72} to estimate the radial distance $r$ from the inferred orbital period $T$:
\begin{eqnarray}
T=310\left[a+(\frac{r}{r_g})^\frac{3}{2}\right]M_7\ (\rm s),\\
r=\left[\frac{T}{310\times M_7} - a\right]^\frac{2}{3} r_g=12^{+8}_{-6}\ r_g, \nonumber
\end{eqnarray}
where $M_7$ is the black hole mass in $10^7$ M$_{\odot}$ units and $a$ is the dimensionless back hole spin. The error bars on the radial distance $r$ are dominated by the uncertainties on the black hole mass.  We note, however, that \citet{vu08} observed that the detection of relativistically redshifted iron K lines could be the result of random fluctuations and \citet{vum16} estimated that at least five periodicities should be sampled to exclude a stochastic process effect.

To enhance the significance of the transient iron K lines, we co-added data extracted from time intervals with similar parameters of the continuum (i.e. column density, 2-10 keV flux and photon index) and in which the flares are most significant. We chose the time intervals peaking at 191, 225, 272, 304 ks, for a total exposure time of 14.6 ks. We will call this co-added spectrum {\it flaring mode} spectrum (Fig. \ref{redadded}, left panel).

\begin{figure}
  \epsfig{file=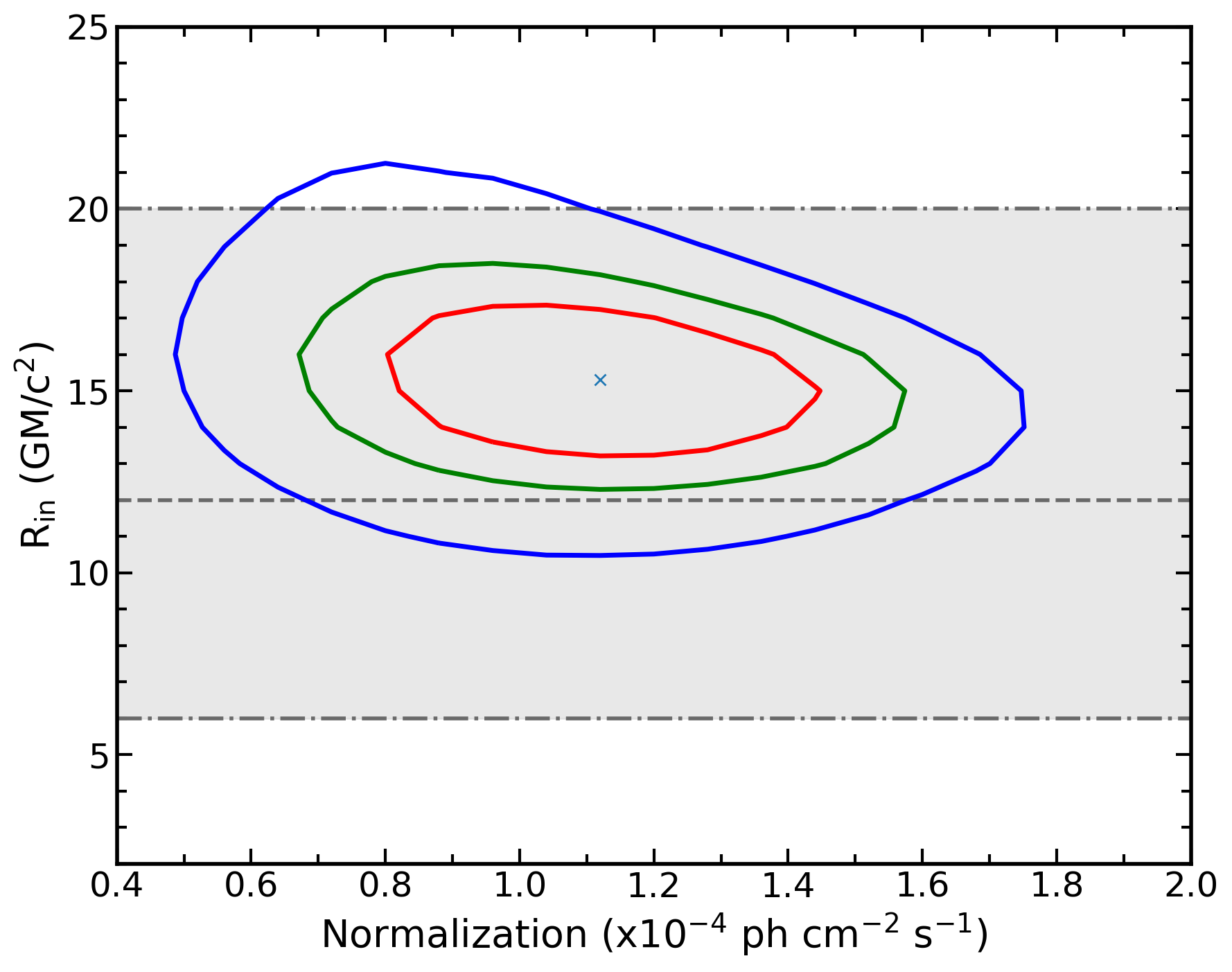, width=0.97\columnwidth}
 \caption{Contour plots between the inner radius of the annular region and the normalization of the {\it red flare} + {\it blue flare II} {\sc KYNrline} component are shown, obtained from the  best fit of the co-added spectra of NGC 2992 in the flaring state. Red, green and blue solid lines indicate 68\%, 90\% and 99\% confidence levels. The dashed  and dotted-dashed black lines indicate the radial distance with its associated errors obtained from the periodicity of the {\it red flare} (see Sect. 3.2 for details). }
  \label{cont}
\end{figure}

When the {\sc KYNrline} model described in the previous section is applied to the {\it flaring mode} spectrum we retrieve a 
best fit $\chi^2/\nu$=135/137, a best fitting energy at rest E$_{\rm R}=6.7\pm0.1$ keV and an inclination angle of the accretion disk $i=40^{\circ}\pm5^{\circ}$ with respect to our line of sight, justifying our previous assumptions on these two parameters. The best fitting values for the {\sc KYNrline} component reproducing the {\it red flare} and {\it blue flare II} are $r_{\rm in}=15\pm3$ r$_g$, $\phi=165^{\circ}-330^{\circ}$ and N=$11.5\pm3.5\times10^{-5}$ ph cm$^{-2}$ s$^{-1}$. This spectral component well reproduces the 5.4 keV and 7.0 keV peaks and its very broad shape (red solid line in the right-panel of Fig. \ref{redadded}) is explained in terms of an emitting region close to a full ring of the accretion disk. The second {\sc KYNrline} component (blue solid line in the right-panel of Fig. \ref{redadded}) is much narrower than the first one and the associated parameters are $r_{\rm in}>50$ r$_g$, $\phi=0^{\circ}-120^{\circ}$ and N=$4.5\pm4.0\times10^{-6}$ ph cm$^{-2}$ s$^{-1}$. The total best fitting model is shown in Fig. \ref{redadded} as a black solid line. \\
\indent We show in Fig. \ref{cont} the contour plots between $r_{\rm in}$ and the normalization: a perfect agreement between $r_{\rm in}$ and the radial distance $r$ retrieved from the periodicity of the {\it red flare} can be seen. Furthermore, the best fit values for the inclination angle of the accretion disk, the inner radius and the normalization of the {\it red flare} + {\it blue flare II} component are perfectly consistent with the ones derived from previous high flux {\it RXTE} and XMM observations \citep{mkt07,sym10}.

\section{Conclusions}
In the previous sections, we presented the analysis of the first simultaneous XMM+{\it NuSTAR} observations of NGC 2992 in an extremely bright state. The source is known to be intrinsically variable, up to a factor of 10 on timescales of days, as observed during the past {\it RXTE} monitoring \citep{mkt07}. It was observed  multiple times by XMM-{\it Newton} between 2010 and 2013, with a 2-10 keV flux always lower than F$_{2-10}<1.7\times 10^{-11}$ erg cm$^{-2}$ s$^{-1}$. The only XMM observation of the source in the 2003 bright state was heavily affected by pile-up \citep{sym10}. Nevertheless, the authors found a relativistic iron K$\alpha$, in accordance with what already observed with {\it RXTE}.  To tackle the high flux levels of the source (i.e. F$_{2-10}>7\times 10^{-11}$ erg cm$^{-2}$ s$^{-1}$), we obtained 60 {\it Swift}-XRT snapshots from March 26, 2019 to December 14, 2019 and the triggered XMM-{\it NuSTAR} observations started on May 7, 2019. We observe a range of fluxes corresponding to accretion rates L$_{\rm Bol}$/L$_{\rm Edd}\simeq$4-6\% and we confirm the physical scenario in which the source exhibits strong redshifted/blueshifted Fe K lines at L$_{\rm Bol}$/L$_{\rm Edd}>$4\% \citep{mkt07, mbb18}. \\
\indent From the XMM-{\it Newton} excess emission map shown in Sect. 3.2, we find hints of a recursive {\it red flare} at $\sim5.4$ keV, with a period $T=41\pm1$ ks. For a maximally rotating black hole spin and considering the black hole mass of the source, the inferred period corresponds to a radial distance $r=12_{-6}^{+8}$ r$_{g\rm }$.  A fully consistent value is found when the {\sc KYNrline} model is applied to the co-added spectra of the flaring states only ($r_{\rm in}=15\pm3$ r$_g$). We find that the {\it red flare} is likely associated to a second spectral component ({\it blue flare II}, at $\sim 7.0$ keV) and can be modelled with a line-emitting annular region close to a full ring of the disk. \\
\indent In the last few years, the technique of mapping the time variability of the flux in excess to the continuum, at different energies, has led to a number of results in bright AGN, with NGC 3516 \citep{imf04}, Mrk 766 \citep{tmg06}, NGC 3783 \citep{tdm07} and Ark 120 \citep{npr16} being the most significant ones. Differently from NGC 3516 and Mrk 766, we cannot constrain a clear evolution of the energy centroid of the emission lines, only changes in fluxes are detected. In our model, the {\it red flare} + {\it blue flare II} component likely arises from large angular regions close to the central black hole ($\phi\simeq150^{\circ}-360^{\circ}$, $r_{\rm in}\simeq10-40$ r$_g$), and well reproduces the excesses observed at 5-5.8 keV and 6.8-7.2 keV.  The width of this sector is not due to the limited spectral quality of our data, but mainly by the necessity to account for both the red and most of the blue excess. On the other hand, the {\it blue flare I} component is much narrower and it is consistent with arising from smaller angular sectors, further away in the disk ($\phi\simeq0^{\circ}-150^{\circ}$, $r_{\rm in}>50$ r$_g$). The hotspot scenario \citep{ruz00,nk01,dov04} is often invoked to explain the observed variability of narrow iron K$\alpha$ emission lines, both in energy and in flux. However, due to the lack of a significant shift in energy of the emission features observed and due to the large angular sectors which best reproduce the red component modulation, the hotspot picture seems unlikely. A single orbital spot cannot reproduce the whole variability patterns and the scenario seems much more complex. In particular, the periodicity of the {\it red flare} which is not observed in the {\it blue flare II} (bottom panels of Fig. \ref{variability}) suggests that their angular distance cannot be resolved and only part of the angular sector is responsible for the periodical excess. Future  X-ray observatories with higher spectral resolution and much larger effective area (such as {\it Athena, eXTP, XRISM}) will be crucial for this kind of studies. \\

\section*{Acknowledgements}
We thank the anonymous referee for her/his suggestions which improved the manuscript. AM thanks A. De Rosa and M. Dov\v{c}iak for useful discussions. AM and RM acknowledge the support of the International Space Science Institute (ISSI Bern, Switzerland). BDM acknowledges support from the European Union's Horizon 2020 research and innovation programme under the Marie Sk{\l}odowska-Curie grant agreement No. 798726 and Polish National Science Center grant OPUS No. 2015/17/B/ST9/03422. We made use of {\sc Astropy},\footnote{\url{http://www.astropy.org}} a community-developed core {\sc Python} package for Astronomy \citep{astropy13, astropy18} and {\sc matplotlib} \citep{h07}. This research has made use of the {\it NuSTAR} Data Analysis Software (NuSTARDAS) jointly developed by the ASI Science Data Center (ASDC, Italy) and the California Institute of Technology (USA).





\bibliographystyle{mnras}
\bibliographystyle{mnras}
\bibliography{sbs} 



\appendix

\section{Best fit results with {\sc KYNrline}}
We present in Table \ref{bestfitPar} the best fit parameters obtained with the {\sc KYNrline} model described in Sect. 3.4.2. We show in Fig. \ref{bfit1} and \ref{bfit2} the spectra, the corresponding best fit models and the relative residuals from the full XMM-{\it Newton} observation.

\begin{table*}
\begin{center}
\begin{tabular}{cccccccccccc}
{\bf Time } & \multicolumn{10}{c}{\bf Best fit parameter} & {\bf $\chi^2/\nu$ }\\
\hline
 & & & & & \multicolumn{3}{c}{\bf Red flare + Blue flare II} & \multicolumn{3}{c}{\bf Blue flare I} &\\
 & N$_{\rm H}$ & $\Gamma$ & E$_{{\rm K}\alpha}$ & N$_{{\rm K}\alpha}$ &r$_{in}$ &$\phi$ & $N_{\rm RS}$ & r$_{in}$ &$\phi$ & $N_{\rm BS}$ & \\
 & $(\times10^{22}$)& & & $(\times10^{-5})$&  &  & $(\times10^{-5})$& & & $(\times10^{-5})$ &\\
   \hline
 \multicolumn{12}{c}{\bf Orbit 1} \\
5 ks&$0.82\pm0.18$ & $1.74\pm0.05$&$6.41\pm0.03$&$8.1\pm2.2$&$20^{+40}_{-10}$&$180^{+160}_{-20}$&$9.0\pm6.2$&-&-&$<$1.7 &114/113 \\
10 ks&$1.08\pm0.18$ & $1.80\pm0.05$&$6.34^{+0.04}_{-0.01}$&$6.3\pm2.3$&$17\pm3$&$140^{+200}_{-15}$&$14.7\pm3.0$&-&-&$<$2.8 &116/115 \\
15 ks&$0.93\pm0.20$ & $1.72\pm0.05$&$6.38\pm0.04$&$7.7\pm2.3$&-&-&$<$7.1&-&-&$<$3.9 &126/117 \\
20 ks&$0.82\pm0.20$ & $1.70\pm0.05$&$6.37^{+0.05}_{-0.03}$&$7.0\pm2.0$&$25^{+15}_{-5}$&$175^{+170}_{-20}$&$9.2\pm6.3$&-&-&$<$4.5 &122/112 \\
25 ks&$0.78\pm0.19$ & $1.65\pm0.05$&$6.38\pm0.02$&$9.0\pm2.5$&-&-&$<$6.7&-&-&$<$3.2 &103/118 \\
30 ks&$1.07\pm0.21$ & $1.74\pm0.05$&$6.37^{+0.06}_{-0.07}$&$2.8\pm2.8$&$45^{+20}_{-10}$&$180^{+190}_{-15}$&$10.7\pm5.5$&$>$80&$100^{+20}_{-30}$&$2.8\pm2.1$ &134/123 \\
35 ks&$1.11\pm0.23$ & $1.71\pm0.06$&$6.31\pm0.07$&$5.5\pm2.5$&-&-&$<$7.0&$30^{+30}_{-10}$&$60^{+80}_{-15}$&$8.2\pm3.0$ &111/115 \\
40 ks&$0.83\pm0.23$ & $1.68\pm0.06$&$6.38\pm0.03$&$7.5\pm2.2$&$45^{+15}_{-10}$&$180^{+200}_{-20}$&$10.5\pm5.0$&-&-&$<1.6$&104/114 \\
45 ks&$0.92\pm0.22$ & $1.68\pm0.06$&$6.38\pm0.04$&$8.7\pm2.2$&$50^{+70}_{-30}$&$160^{+200}_{-160}$&$6.7\pm4.3$&-&-&$<1.6$&104/113 \\
50 ks&$0.87\pm0.19$ & $1.66\pm0.05$&$6.37\pm0.03$&$7.6\pm2.0$&-&-&$<$4.6&-&-&$<2.6$&115/114 \\
55 ks&$1.00\pm0.21$ & $1.66\pm0.05$&$6.38\pm0.03$&$9.8\pm2.0$&-&-&$<$4.4&-&-&$<3.2$&127/114 \\
60 ks&$0.90\pm0.19$ & $1.69\pm0.05$&$6.38\pm0.05$&$6.3\pm2.9$&-&-&$<$8.5&$140^{+30}_{-90}$&$30^{+60}_{-100}$&$2.8\pm2.4$&113/115 \\
65 ks&$0.91\pm0.18$ & $1.72\pm0.05$&$6.38\pm0.03$&$8.1\pm2.4$&-&-&$<$10.2&-&-&$<4.4$&127/117 \\
70 ks&$0.90\pm0.18$ & $1.66\pm0.05$&$6.39\pm0.05$&$6.1\pm2.3$&-&-&$<$9.5&-&-&$<4.5$&127/117 \\
75 ks&$0.95\pm0.18$ & $1.72\pm0.05$&$6.41\pm0.03$&$9.5\pm2.5$&-&-&$<$10.3&$180^{+20}_{-170}$&$50^{+60}_{-15}$&$2.7\pm2.3$&123/111 \\
80 ks&$0.90\pm0.20$ & $1.66\pm0.05$&$6.40\pm0.03$&$9.0\pm2.5$&-&-&$<$5.3&-&-&$<$5.0&117/114 \\
85 ks&$0.81\pm0.18$ & $1.69\pm0.05$&$6.38\pm0.05$&$5.7\pm2.4$&$40^{+10}_{-5}$&$175^{+155}_{-15}$ & $12.0\pm5.3$&-&-&$<$2.5&127/114 \\
90 ks&$0.85\pm0.18$ & $1.68\pm0.05$&$6.40\pm0.07$&$5.3\pm3.0$&$80^{+70}_{-35}$&$180^{+170}_{-20}$&$9.9\pm5.0$&- & -& $<$4.2&118/120 \\
95 ks&$0.90\pm0.19$ & $1.69\pm0.06$&$6.41\pm0.04$&$6.4\pm2.1$&- & -& $<$6.0&-&-&$<$2.7&100/119 \\
100 ks&$1.15\pm0.20$ & $1.74\pm0.05$&$6.38\pm0.03$&$8.1\pm2.2$&- & -& $<$7.7&-&-&$<$3.5&121/117 \\
105 ks&$0.94\pm0.20$ & $1.67\pm0.06$&$6.38\pm0.03$&$9.3\pm2.3$&- & -& $<$6.7&-&-&$<$4.4&121/116 \\
110 ks&$0.81\pm0.20$ & $1.67\pm0.06$&$6.40\pm0.03$&$8.4\pm2.3$&- & -& $<$11.5&-&-&$<$3.8&102/115 \\
115 ks&$0.77\pm0.21$ & $1.60\pm0.06$&$6.39\pm0.03$&$8.7\pm2.3$&- & -& $<$8.2&-&-&$<$4.0&114/114 \\
120 ks&$0.76\pm0.20$ & $1.62\pm0.06$&$6.38\pm0.04$&$6.0\pm2.5$&$35^{+25}_{-15}$ & $180^{+200}_{-20}$& $9.2\pm5.6$&$150^{+50}_{-120}$&$60^{+70}_{-40}$&$2.8\pm2.3$&129/111 \\
125 ks&$0.80\pm0.21$ & $1.60\pm0.06$&$6.38\pm0.04$&$6.2\pm2.5$&$30^{+5}_{-10}$ & $180^{+300}_{-30}$& $9.4\pm5.1$&-&-&$<$4.2&132/114 \\
130 ks&$1.00\pm0.20$ & $1.71\pm0.06$&$6.38\pm0.04$&$6.2\pm2.5$&$10^{+5}_{-3}$ & $180^{+200}_{-20}$& $16.0\pm8.4$&-&-&$<2.9$&124/114 \\
 \hline
 \multicolumn{12}{c}{\bf Orbit 2} \\
177 ks&$0.81\pm0.23$ & $1.63^{+0.07}_{-0.06}$&$6.40\pm0.03$&$8.7\pm2.0$&-&-&$<$7.6&-&-&$<$1.2 &108/110 \\
182 ks& $0.92\pm0.21$& $1.64\pm0.06$&$6.38^{+0.04}_{-0.03}$&$6.1\pm2.0$&-&-&$<$5.7&$>$150&$30^{+100}_{-40}$&$2.6\pm2.0$&127/111\\
187 ks&$0.98^{+0.20}_{-0.32}$ & $1.67^{+0.08}_{-0.09}$&$6.33^{+0.05}_{-0.03}$&$7.3\pm2.2$&$43^{+15}_{-32}$&$170^{+250}_{-40}$&$7.7\pm4.5$&$>45$&$70^{+50}_{-10}$&$7.7\pm2.5$ & 127/108\\
191 ks&$0.86\pm0.25$ &$1.63^{+0.06}_{-0.05}$ &$6.31\pm0.05$&$5.0\pm2.1$&$14^{+5}_{-7}$&$180^{+180}_{-15}$&$11.5\pm4.7$&$>400$&$100^{+240}_{-50}$&$4.7\pm2.0$&112/110\\
196 ks&$0.78\pm0.25$ & $1.59\pm0.06$&$6.37\pm0.03$&$7.9\pm1.8$&-&-&$<$9.6&$>350$&$80^{+250}_{-25}$&$5.4\pm2.2$&94/112\\
202 ks& $0.98\pm0.20$& $1.65\pm0.05$&$6.36\pm0.03$&$7.4^{+1.8}_{-2.0}$&$28_{-8}^{+10}$&$120^{+280}_{-30}$&$8.4\pm2.5$&-&-&$<$2.5 & 130/111\\
208 ks& $0.81\pm0.20$& $1.59\pm0.06$&$6.39\pm0.05$&$7.6\pm2.5$&-&-&$<$6.9&-&-&$<$4.6& 110/115\\
214 ks& $1.03\pm0.22$& $1.67\pm0.07$& $6.37\pm0.05$&$5.5\pm2.5$ &$70^{+70}_{-45}$ & $180^{+250}_{-30}$ &$6.2\pm3.2$&$50^{+200}_{-35}$ &$65^{+150}_{-25}$&$3.2\pm2.7$ & 85/109\\
220 ks& $0.86\pm0.22$& $1.57\pm0.06$& $6.41\pm0.03$&$7.5\pm2.0$ &- & - &$<8.2$&- &-&$<2.9$ &103/113\\
225 ks& $1.02\pm0.20$&$1.65\pm0.05$ &$6.35\pm0.05$ & $6.0\pm2.0$&$18^{+5}_{-3}$&$160^{+190}_{-15}$&$11.6\pm5.1$&-&-&$<$2.3& 107/127\\
231 ks&$0.86\pm0.21$ & $1.62\pm0.06$& $6.39\pm0.04$& $7.1\pm1.9$&$20^{+25}_{-15}$&$165^{+180}_{-20}$&$6.0\pm4.8$&-&-&$<$3.1&133/111\\
237 ks& $0.73\pm0.20$& $1.60\pm0.06$& $6.41\pm0.05$&$5.8\pm2.4$ &-&-&$<$3.6&$20^{+30}_{-5}$&$45^{+25}_{-15}$&$6.1^{+3.0}_{-4.1}$&114/110\\
243 ks&$0.89\pm0.21$ & $1.62\pm0.06$&$6.38\pm0.04$&$7.2\pm2.3$&-&-&$<$9.0&-&-&$<$6.0&142/113\\
249 ks&$1.00\pm0.21$ & $1.65\pm0.06$&$6.39\pm0.03$&$7.3\pm2.0$&-&-&$<$9.5&$>50$&$60^{+180}_{-60}$&$2.6\pm2.2$&114/111\\
254 ks&$0.80\pm0.20$ & $1.63\pm0.07$ &$6.39\pm0.03$&$6.0\pm2.0$&$30^{+15}_{-10}$&$130^{+220}_{-40}$&$4.8\pm3.4$&-&-&$<$3.6 & 98/113\\
260 ks&$0.81\pm0.19$ & $1.59\pm0.06$&$6.39\pm0.03$&$8.8\pm2.1$&-&-&$<$9.0&-&-&$<$3.5&113/118\\
266 ks& $0.78\pm0.21$&$1.62\pm0.06$ &$6.36\pm0.04$&$6.2\pm2.0$&$17^{+8}_{-10}$&$160^{+180}_{-40}$&$6.2\pm5.2$&-&-&$<$2.5& 117/112\\
272 ks&$0.86\pm0.21$&$1.61\pm0.06$ &$6.42\pm0.04$&$8.9\pm2.0$&$15^{+4}_{-6}$&$160^{+200}_{-15}$&$12.9\pm5.8$&-&-&$<$2.0 & 85/113\\
278 ks& $0.80\pm0.18$&$1.56\pm0.06$  &$6.40\pm0.02$&$7.5\pm2.0$&-&-&$<$7.2&-&-&$<$2.5&106/116\\
283 ks& $0.88\pm0.20$&$1.68\pm0.05$ &$6.39\pm0.03$&$9.0\pm2.0$&-&-&$<$7.2&-&-&$<$2.5&121/116\\
289 ks&$0.73\pm0.20$ &$1.60\pm0.05$ &$6.38\pm0.02$&$8.0\pm2.0$&-&-&$<$9.8&$>180$&$20^{+50}_{-80}$& $2.2\pm2.0$& 119/111\\
295 ks& $0.91\pm0.20$&$1.60\pm0.06$ &$6.40\pm0.03$&$8.0\pm2.0$&-&-&$<$8.8&-&-&$<$3.0&121/116\\
301 ks&$0.89\pm0.20$ & $1.61\pm0.06$&$6.36\pm0.04$ &$5.8\pm2.1$ &-&-&$<$9.2&$30^{+120}_{-15}$&$40^{+70}_{-10}$&$4.4\pm3.2$&87/112\\
304 ks&$1.06\pm0.30$ & $1.68\pm0.08$&$6.43\pm0.05$&$6.9\pm2.7$&$15^{+5}_{-5}$&$130^{+270}_{-40}$&$11.7\pm8.0$&$>200$&$50^{+100}_{-40}$&$3.2\pm2.5$&102/102\\
\hline
\end{tabular}
\end{center}
\caption{\label{bestfitPar} Best fit parameters of the time-resolved XMM analysis. Column densities are in cm$^{-2}$ units, energies are in keV units, normalizations are in ph cm$^{-2}$ s$^{-1}$ units, r$_{\rm in}$ are in r$_g$ units and $\phi$ are in degrees units. See Sect. 3.3.2 for details on the model.}
\end{table*}

\begin{figure*}
 \epsfig{file=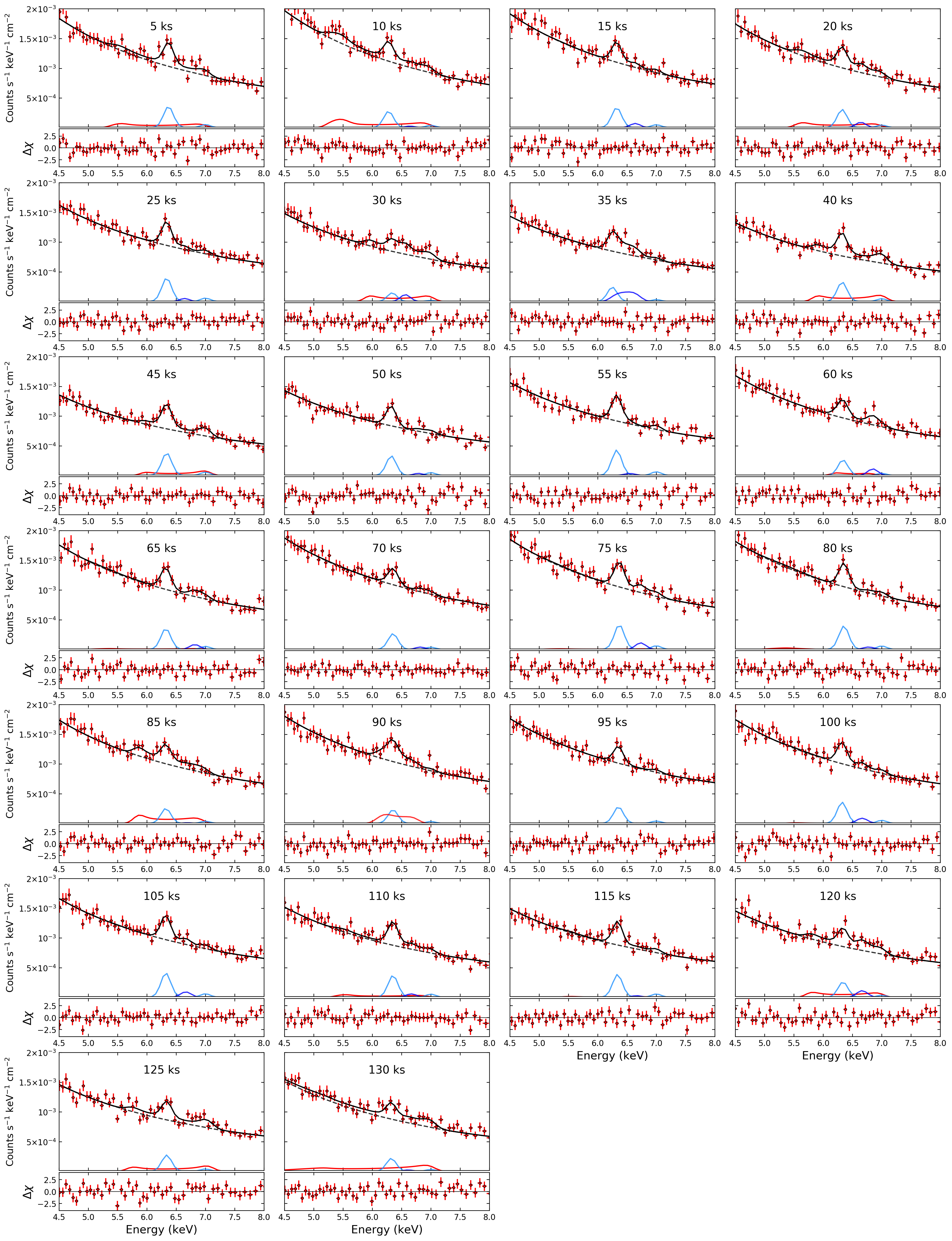, width=2.\columnwidth}
 \caption{XMM EPIC pn data (divided by the instrumental effective area), best fits and residuals are shown, in the 4.5-8.0 keV band. Black solid lines indicate best fit models, dashed black lines indicate the power law spectral component, cyan solid lines indicate best fit iron K$\alpha$ and K$\beta$ components, red and blue solid lines indicate the two {\sc KYNrline} components ({\it red flare} + {\it blue flare  II } and {\it blue flare  I }).}
  \label{bfit1}
\end{figure*}

\begin{figure*}
 \epsfig{file=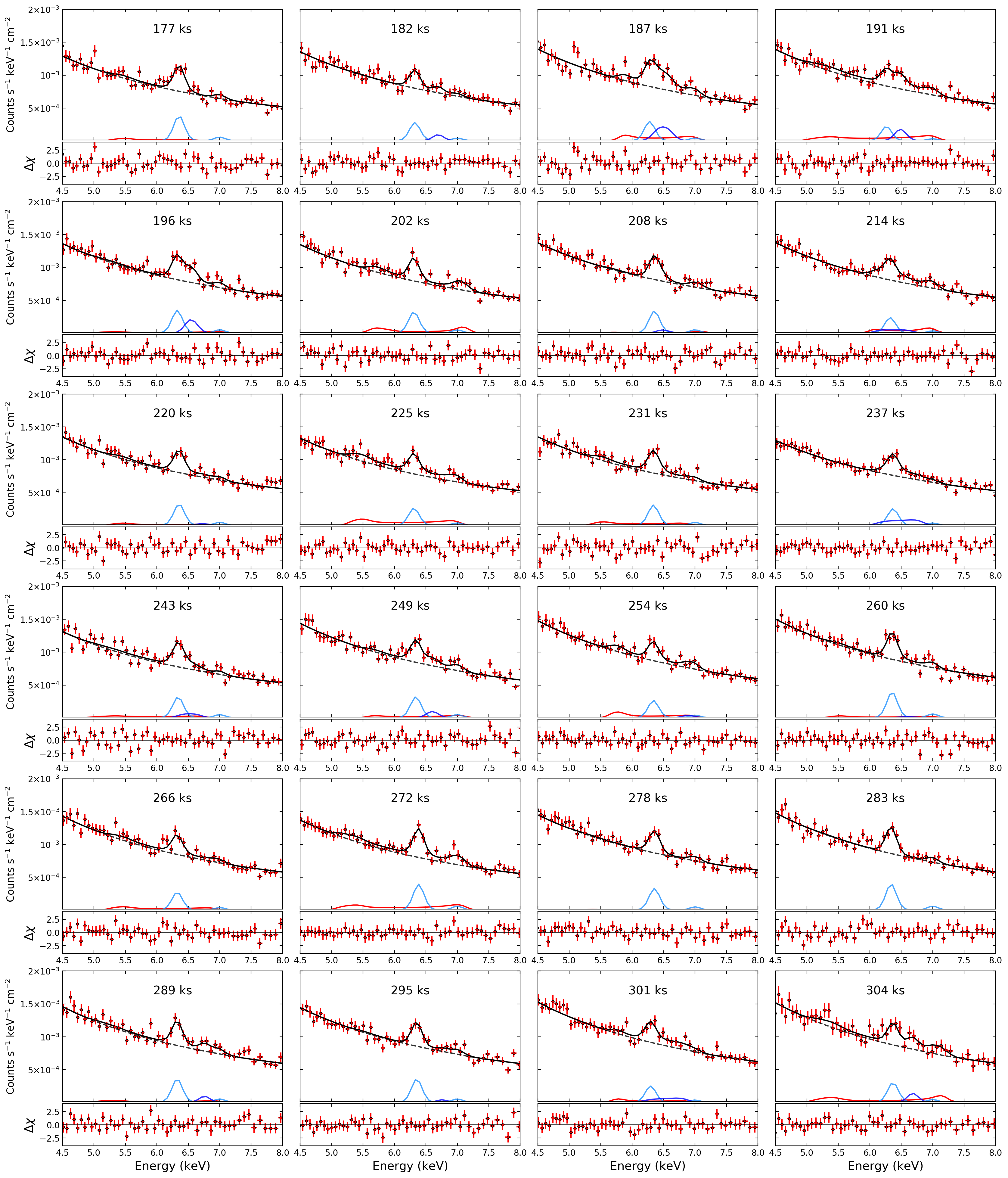, width=2.\columnwidth}
 \caption{{\it continued.}}
  \label{bfit2}
\end{figure*}


\bsp	
\label{lastpage}
\end{document}